\documentclass[12pt,prd,amsmath,amssymb,nofootinbib,tightenlines,floatfix,preprintnumbers]{revtex4}

\usepackage{epsfig,latexsym,cancel,amssymb,amsmath,mathrsfs,verbatim}
\usepackage{color}
\usepackage{graphicx}

\newcommand{\be}{\begin{equation}}
\newcommand{\ee}{\end{equation}}
\newcommand{\bea}{\begin{eqnarray}}
\newcommand{\eea}{\end{eqnarray}}
\newcommand{\ba}{\begin{array}}
\newcommand{\ea}{\end{array}}

\newcommand{\Lint}{{\mathscr{L}_{\textrm{int}}}}

\long\def\symbolfootnote[#1]#2{\begingroup%
\def\thefootnote{\fnsymbol{footnote}}\footnote[#1]{#2}\endgroup} 

\newcommand{\eq}[1]{Eq.~\eqref{#1}}
\newcommand{\eqs}[2]{Eqs.~\eqref{#1} and \eqref{#2}}

\renewcommand{\sec}[1]{Sec.~\ref{#1}}
\newcommand{\fig}[1]{Fig.~\ref{#1}}

\newcommand{\vect}[1]{\mathbf{#1}}
\newcommand{\abs}[1]{\left\lvert #1\right\rvert}
\newcommand{\grad}{\mbox{\boldmath$\nabla$}}

\newcommand{\pd}[2]{\frac{\partial #1}{\partial #2}}

\newcommand{\GeV}{\text{ GeV}}
\newcommand{\eosc}{\epsilon_{\text{osc}}}
\newcommand{\ewall}{\epsilon_{\text{wall}}}
\newcommand{\ecoll}{\epsilon_{\text{coll}}}

\DeclareMathOperator{\diag}{diag}

\begin{document}

\preprint{MIT-CTP 4269}

\title{Resonant Flavor Oscillations in Electroweak Baryogenesis}

\author{Vincenzo Cirigliano}
\affiliation{Theoretical Division, Los Alamos National Laboratory, Los Alamos, NM, 87545, USA}
\author{Christopher Lee} 
\affiliation{Center for Theoretical Physics,   \\ Massachusetts Institute of Technology, Cambridge, MA, 02139, USA}
\author{Sean Tulin} 
\affiliation{Theory Group, TRIUMF, 4004 Wesbrook Mall, Vancouver, BC, V6T 2A3, Canada}

\date{\today}

\begin{abstract}

Electroweak baryogenesis (EWBG) in extensions of the Standard Model will be tested quantitatively in upcoming nuclear and particle physics experiments, but only to the extent that theoretical computations are robust.  Currently there exist orders-of-magnitude discrepancies between treatments of charge transport dynamics during EWBG performed by different groups, each relying on different sets of approximations.  In this work, we introduce a consistent power counting scheme (in ratios of length scales) for treating systematically the dynamics of EWBG: CP-asymmetric flavor oscillations, collisions, and diffusion.  Within the context of a simplified model of EWBG, we derive the relevant Boltzmann equations using non-equilibrium field theory, and solve them exactly without ansatz for the functional form of the density matrices. We demonstrate the existence of a resonant enhancement in charge production when the flavor oscillation length is comparable to the wall thickness. 
We compare our results with the existing treatment of EWBG by Konstandin, Prokopec, Schmidt, and Seco (KPSS) who previously identified the importance of flavor oscillations in EWBG.  We conclude: (i) the power counting of KPSS breaks down in the resonant regime, and (ii) this leads to substantial underestimation of the charge generated in the unbroken phase, and potentially of the final baryon asymmetry.  

\end{abstract}

\pacs{}
\maketitle

\newpage

\section{Introduction}
\label{sec:intro}

Electroweak baryogenesis (EWBG) is an attractive mechanism to explain the origin of the baryon asymmetry of the Universe (BAU).  Because the relevant physics is accessible at the electroweak scale, nuclear and particle physics experiments can directly test two of the three Sakharov conditions~\cite{Sakharov:1967dj}.  The departure from equilibrium during the electroweak phase transition (EWPT) can be addressed in collider studies (e.g.,~\cite{Profumo:2007wc}), while CP violation can be probed through low-energy observables such as 
electric dipole moments (EDM)~\cite{Cirigliano:2009yd}.

In EWBG, electroweak symmetry breaking proceeds through a first-order EWPT, where bubbles of broken SU(2)$_L\times$U(1)$_Y$ symmetry nucleate and expand in a background of unbroken symmetry.  CP-violating interactions within the bubble wall produce a CP-asymmetric charge density of left-handed fermions.  This charge, diffusing ahead of the bubble wall into the unbroken phase, is converted into a baryon asymmetry through non-perturbative electroweak sphaleron processes~\cite{Kuzmin:1985mm}.  Lastly, the baryon asymmetry is captured by the advancing bubble wall and freezes out.  The Sakharov conditions are satisfied provided (i) the EWPT is ``strongly'' first-order, otherwise electroweak sphalerons are too active in the broken phase and the baryon density is washed out, and (ii) there exists sufficient CP violation to generate  the observed BAU.  Although neither condition is met in the Standard Model (SM)~\cite{Farrar:1993sp}, EWBG may be viable in the Minimal Supersymmetric SM (MSSM)~\cite{Cirigliano:2009yd, Carena:2008vj} and other scenarios 
beyond the SM~\cite{bsm}.

Ultimately, the final baryon asymmetry in EWBG is roughly proportional to the total CP-asymmetric charge that is generated and transported, by diffusion, into the unbroken phase where weak sphalerons are active.  However, within a given model, this charge transport computation is a complicated problem: one must solve a network of Boltzmann equations governing the generation, diffusion, and equilibration of charges in the vicinity of the moving bubble wall.  
To date, there exists no treatment of EWBG dynamics that includes all of these aspects in a consistent and theoretically robust framework.  And yet, such a treatment is crucial for making a quantitative connection between EWBG and experiment.  Here, a key insight was provided by Refs.~\cite{konstandin,Prokopec:2003pj} in recognizing the importance of CP-violating flavor oscillations.  These oscillations arise through spacetime-dependent flavor mixing induced by the varying background Higgs field.

In the thick bubble wall regime ($L_w \gg T^{-1}$), 
 flavor oscillations are formally the leading source of CP violation in a gradient expansion in powers of  $(L_w T)^{-1}$, arising at linear order (e.g., in the MSSM, $L_w \sim 20/T$~\cite{Moreno:1998bq}).  
At order $(L_w T)^{-2}$, one finds an additional CP-violating source from the spin-dependent ``semi-classical force"~\cite{Cline:1997vk, Cline:2000nw, Kainulainen:2001cn,Kainulainen:2002th,Zhou:2008yd}.
Aside from Refs.~\cite{konstandin,Prokopec:2003pj}, all previous EWBG computations have swept flavor oscillations under the rug: quantum coherence between states was neglected, and this CP-violating source 
was treated perturbatively 
in terms of scattering off the external background Higgs field or its gradient.
 It is unknown whether or not such prescriptions are justified.  Furthermore, Ref.~\cite{konstandin} found that 
 CP-violating charge  densities generated by flavor mixing are  localized to the bubble wall and do {\it not} diffuse into the unbroken phase, in contrast to other treatments.  
 On the other hand, if significant charge diffusion occurs  it would lead to a substantial enhancement of baryon number generation by transporting charge outside the bubble where sphalerons are active~\cite{Cohen:1994ss,Joyce:1994zn,Huet:1995sh}.  Therefore, this issue is highly relevant for experimental tests of EWBG.  Indeed, according to Ref.~\cite{konstandin}, EWBG in the MSSM is ruled out due to null electron and neutron electric dipole moment (EDM) searches, while it is still viable (to varying degrees) according to other EWBG computations~\cite{Cirigliano:2009yd,Carena:2008vj}.

It is clear that quantitative EWBG computations require a consistent analysis of CP-violating flavor oscillations, collisions in the plasma, and diffusion, 
which is still missing in the literature.             
In Ref.~\cite{Cirigliano:2009yt}, we performed a first step 
towards such a complete treatment: we studied a model of two 
scalars fields  $\Phi_{L,R}$ that mix through 
a purely {\it time}-dependent CP-violating mass matrix and have 
flavor-sensitive interactions  with a thermal bath of  scalar bosons ($A$) in equilibrium.
We derived using non-equilibrium field theory~\cite{SchwingerKeldysh} quantum Boltzmann equations for the two-flavor density matrices (Wigner functions), relying on a systematic expansion scheme in ratios of time scales.  Our work demonstrated --- for the first time in a consistent framework ---  the resonant baryogenesis regime discussed in 
Refs.~\cite{Carena:1997gx,Riotto:1998zb,Carena:2000id,Carena:2002ss,Balazs:2004ae,Lee:2004we,Cirigliano:2006wh} and placed it on a more rigorous theoretical footing.  Importantly, we showed that the resonance occurs when the flavor oscillation period is comparable to the variation scale of the background Higgs field.

In this work, we generalize the analysis of Ref.~\cite{Cirigliano:2009yt} to account for a spacetime-dependent background field geometry, as in a moving bubble wall.  
Non-homogeneity of the bubble geometry is a key ingredient of EWBG, essential for generation of a CP-violating charge that undergoes subsequent transport by diffusion.  We work within the context of the same toy model of Ref.~\cite{Cirigliano:2009yt}.
Our work is organized as follows:

\begin{itemize}

\item In Sec.~\ref{sec:toy}, we present our toy model for two mixing scalars ($\Phi_{L,R}$) with a spacetime-dependent mass matrix induced by the bubble wall.  We discuss how this model maps onto more ``realistic'' baryogenesis models.

\item We derive the Boltzmann equations for the occupation numbers and quantum coherence of $\Phi_{L,R}$ states in Sec.~\ref{Sect:QBE}.  We present both a heuristic, intuitive derivation, and a more rigorous one using non-equilibrium  field theory.  We also discuss the necessary conditions for CP violation.

\item 
Using numerical methods,
we solve the Boltzmann equations exactly in Sec.~\ref{Sect:solution}, without ansatz for  the form of the density matrices.  Our results clearly demonstrate the existence of charge diffusion, the role of flavor oscillations in generating CP asymmetries, and how charge generation is enhanced in the resonant regime $|m_1 - m_2| \lesssim 10 L_w^{-1}$, where $m_{1,2}$ are the mass eigenvalues of the two-scalar system.  Here, we also provide a useful analogy with spin precession in a varying magnetic field.
 
\item In Sec.~\ref{Sect:comparison}, we highlight the differences between our approach and that of Refs.~\cite{konstandin,Prokopec:2003pj}.  
Our major disagreement stems from Refs.~\cite{konstandin,Prokopec:2003pj} power counting the solutions to the Boltzmann equations in powers of $(L_w T)^{-1}$, while we do not.  We demonstrate that this power-counting argument breaks down in the resonant regime, effectively negates the possibility of diffusion, and substantially underestimates the amount of charge generated during the EWPT.

\end{itemize}

The virtue of working within our toy model is that it can be solved both exactly (with numerical techniques), as well as within 
various  approximation schemes. 
This offers the possibility to study 
some of the key assumptions used in current approaches 
and to quantify the attendant uncertainties.  In this work we have focused  on the comparison with what has so far been considered the 
state-of-the-art calculation in Ref.~\cite{konstandin}.  
In forthcoming work, we will study in detail  
the diffusion approximation, invoked in essentially all EWBG calculations to make  
the problem tractable.  In future work we will also extend the application of our methods to fermions, which are an essential ingredient in EWBG since sphalerons couple only to fermions.


\section{Baryogenesis Toy Model}
\label{sec:toy}

Two-flavor dynamics of scalars are highly relevant in extensions of the MSSM, where top squarks $(\widetilde t_L, \widetilde t_R)$ may account for the BAU~\cite{Blum:2010by}.  These models are necessarily complicated by their large number of degrees of freedom ($g_* \sim 200$) and many different types of interactions.  Here, we consider a much simplified model: a two-flavor scalar system, with fields $\Phi\equiv(\Phi_{L},\Phi_R)$, described by the Lagrangian
\be
\mathscr{L} = \partial_\mu \Phi^\dagger \, \partial^\mu \Phi - \Phi^\dagger \, M^2 \, \Phi + \Lint  \label{eq:model} \; .
\ee
The key ingredients of our model are:
\begin{itemize}
\item The mass matrix $M^2(x)$ is spacetime-dependent, assumed to a function of the varying background Higgs field(s) associated with the expanding bubble.  The variation of $M^2$ across the phase boundary generates $\Phi_{L,R}$ charge through CP-violating coherent flavor oscillations.
\item $\Lint$ describes the interactions of $\Phi$ with the remaining degrees of freedom in the plasma.  These collisions govern the damping of flavor oscillations and the effective diffusion of these charges in the plasma.  
\end{itemize} 
We model the plasma during the time of the EWPT as a thermal bath of real scalar bosons $A$, assumed to be in equilibrium at temperature $T$, coupled to $\Phi$ via
\be
\Lint = \, - \, \frac{1}{2} \: A^2 \, \Phi^\dagger \, y \, \Phi  \label{eq:lint} \; , \qquad y=\left( \ba{cc} y_L & 0 \\ 0  & y_R \ea \right) \; .
\ee
We take the matrix of coupling constants $y$ to be  diagonal; this defines the basis of flavor eigenstate fields $\Phi_{L,R}$ (``flavor basis'').

The mass matrix can be parametrized as
\be
M^2(x) = \left( \ba{cc} m_L^2(x) & v(x) \, e^{-i  \alpha(x)} \\ v(x) \, e^{i  \alpha(x)} & m_R^2(x) \ea \right) \; .  
\ee
It is convenient to transform Eq.~\eqref{eq:model} into the basis of local mass eigenstates (``mass basis'').  We
diagonalize the mass matrix with the spacetime-dependent transformation matrix $U(x)$, such that
\label{eq:diagonalization}
\bea
m^2(x) \equiv  \left( \ba{cc} m^2_1(x) & 0 \\ 0 & m_2^2(x) \ea \right) = U^\dagger M^2  U    \; , \quad 
U(x) = \begin{pmatrix} \cos\theta(x) & -\sin\theta(x)\, e^{-i \alpha(x)} \\ \sin\theta(x)\, e^{i \alpha(x)} & \cos\theta(x) \end{pmatrix}  
\eea
with  
\begin{align}
m^2_{1,2} = \frac{1}{2}  (m_L^2 + m_R^2)   \pm \frac{1}{2}  {\rm sign} (m_L^2 - m_R^2)  \sqrt{(m_L^2 - m_R^2)^2  + 4 \, v^2 } \; ,
\ 
\tan (2\theta) =  \frac{2 \, v}{m_L^2 - m_R^2}.
\label{eq:params}
\end{align}
This diagonalization defines the mass basis fields $\phi \equiv (\phi_1, \, \phi_2) \equiv U^\dagger \Phi$.
The Lagrangian, in the mass basis, is
\be
\label{eq:massbasis} 
\mathscr{L} = \partial_\mu \phi^{\dagger} \, \partial^\mu \phi - \phi^{\dagger} m^2  \phi   - \phi^\dagger \, \Sigma^\mu  \partial_\mu \phi + \partial_\mu \phi^\dagger \, \Sigma^\mu  \phi - \phi^\dagger \, \Sigma^\mu \Sigma_\mu  \phi + \Lint \;, 
\ee
where 
\be \label{eq:sigma}
\Sigma^\mu(x) \equiv U^\dagger(x) \, \partial^\mu U(x) = 
\begin{pmatrix}
0 & -e^{-i\alpha} \\
e^{i\alpha} & 0
\end{pmatrix}
\partial^\mu \theta 
+
\begin{pmatrix}
i\sin^2\theta & \frac{i}{2}\sin 2\theta e^{-i\alpha} \\
\frac{i}{2}\sin 2\theta e^{i\alpha} & -i\sin^2\theta
\end{pmatrix}
\partial^\mu \alpha \, .
\ee
In this basis, the interaction becomes
\be
\Lint = \, - \, \frac{1}{2} \: A^2 \, \phi^\dagger \, Y \, \phi \;, \qquad Y(x) \equiv U^\dagger(x) \, y \, U(x) \;.  
\ee

During the EWPT, the background Higgs field has a bubble geometry.  We assume a spherical bubble expanding in the $\hat{\mathbf r}$ direction, with wall thickness $L_{w} \gg T^{-1}$ and velocity $v_w \ll 1$. Typically, in the MSSM, one finds $L_w\sim 20/T$ \cite{Moreno:1998bq} and $v_w\sim 0.05$ \cite{John:2000zq}.  At late time (compared to the nucleation time), the bubble profile can be approximated as planar 
and physical quantities depend only on the coordinate $z \equiv (r-v_w t)$, the distance to the wall.  Motivated by realistic bubble wall profiles~\cite{Moreno:1998bq}, we take
\bea
\label{eq:kink}
v(z) = \frac{v_0}{2} \left( \, 1- \tanh\frac{2z}{L_{w}} \, \right) \, , \quad \alpha(z) = \frac{\alpha_0}{2} \left( \, 1- \tanh\frac{2z}{L_{w}} \, \right) \,  . 
\eea
The $z<0$ $(z>0)$ region corresponds to the (un)broken phase.
Additionally, for simplicity we take constant diagonal elements $m_{L,R}^2$.


\section{ Quantum  Boltzmann Equations}
\label{Sect:QBE}

\subsection{Kinetic Theory Derivation}

Kinetic theory, described by Boltzmann equations, is a useful tool to describe the dynamics of an ensemble of quantum states \cite{Calzetta:1986cq,Blaizot:1992gn,Arnold:2002zm,Arnold:2003zc}.  For a single species, 
characterized by the distribution function $f(\mathbf k,x)$ of states with momentum $k^\mu \! = \! (\omega_{\mathbf k}, \mathbf k)$ and spacetime coordinate $x^\mu \equiv (t,\mathbf x)$,
the usual Boltzmann equation is (in flat spacetime)
\be
(\partial_t + \mathbf{v} \cdot \nabla_{\mathbf x} + \mathbf{F} \cdot \nabla_{\mathbf k} ) \,f(\mathbf k,x)  = \mathscr{C}(\mathbf k,x) \; .  \label{eq:usual}
\ee
Here, $\mathbf v \equiv \mathbf k/\omega_{\mathbf k}$ is the velocity and $\mathbf{F}(\mathbf k,x)$ is the force associated with the variation of an external potential over length scale $L_{\textrm{ext}}$.  The collision term $\mathscr{C}(\mathbf k,x)$ characterizes scattering interactions, with mean free path $L_{\textrm{mfp}}$.  
The Boltzmann picture is valid only in the semi-classical limit, where $L_{\textrm{mfp}}, \, L_{\textrm{ext}} \gg  L_{\textrm{int}}$, where $L_{\textrm{int}} \! = \! |\mathbf k|^{-1}$ is the ``intrinsic'' de Broglie wavelength\footnote{
This statement follows from the uncertainty principle: $\Delta k \, \Delta x \gtrsim 1$.
By describing the system in terms of a distribution $f$, it is assumed that states have well-defined momenta, such that $\Delta k \ll |\mathbf k|$.  At the same time, it is assumed that the force term acts locally at $\mathbf x$, while the collision term is formulated in terms of localized, single scattering interactions.  Both assumptions require that states are sufficiently localized in position with respect to the relevant scales: $\Delta x \ll L_{\textrm{mfp}}, L_{\textrm{ext}}$.  Thus, $L_{\textrm{mfp}}, \, L_{\textrm{ext}} \gg \Delta x \gtrsim 1/\Delta k \gg  L_{\textrm{int}}$.}.

In the two-flavor case, a new effect can arise: flavor oscillations due to quantum coherence between different mass eigenstates.  The relevance of flavor oscillations depends on the oscillation length scale 
$L_{\textrm{osc}} \propto  1/(\omega_1 - \omega_2) \sim  |\mathbf k|/\Delta m^2$, 
where $\Delta m^2\! = \! m_1^2-m_2^2$.  Clearly, if we want to include quantum coherence in our dynamics, Eq.~\eqref{eq:usual} must be generalized. 

In the context of EWBG, the relevant length scales are given as follows:
\begin{itemize}
\item The typical de Broglie wavelength is $L_{\textrm{int}} = |\mathbf k|^{-1} \sim T^{-1}$.
\item The external length scale $L_{\textrm{ext}}$ is set by the wall thickness $L_{w}$.  Previous studies have found $L_{w} \sim (20-30) T^{-1}$ in the MSSM~\cite{Moreno:1998bq} and $L_{w} \sim (2-40) T^{-1}$ in extensions of the MSSM~\cite{Blum:2010by,Huber:2000mg}.
\item The oscillation length $L_{\textrm{osc}}$ is determined by the $\Phi$ mass spectrum. 
In the thick wall regime, CP asymmetries are maximized for $L_{w} \sim L_{\textrm{osc}}$ 
(see Ref.~\cite{Cirigliano:2009yt} and the discussion in  Sec.~\ref{Sect:solution}); thus, the $L_{\textrm{osc}} \gg L_{\textrm{int}}$ case is the most interesting for EWBG.
\item The mean free path satisfies $L_{\textrm{mfp}} \gg L_{\textrm{int}}$ if the couplings $y_{L,R}$ are perturbative.
\end{itemize}
Therefore, we assume in our analysis that the following ratios are small parameters:
\be
\ewall \equiv \frac{L_{\textrm{int}}}{L_{w}} \; , \quad 
\ecoll \equiv \frac{L_{\textrm{int}}}{L_{\textrm{mfp}}} \; , \quad
\eosc \equiv \frac{L_{\textrm{int}}}{L_{\textrm{osc}}} \;  ,
\ee
collectively denoted as $\epsilon$.  

In the $\epsilon \ll 1$ limit, a Boltzmann-like description of a multi-flavor system is still possible, despite the inherently quantum nature of the coherence between states.  In this case, one must promote $f(\mathbf k, x)$ to a density matrix: the diagonal elements denote occupation numbers of states, while the off-diagonal elements describe coherence between those states. The Boltzmann equation becomes a matrix equation:
\be
(\partial_t + \mathbf{v} \cdot \nabla_{\mathbf x} + \mathbf{F} \cdot \nabla_{\mathbf k} ) \, f(\mathbf k,x) = - i \left[ \Omega_{\mathbf k}, \, f(\mathbf k,x) \right] + \mathscr{C}[f,\bar f](\mathbf k,x) \; .  \label{eq:flav}
\ee
The general structure is nearly identical to Eq.~\eqref{eq:usual}, except for two important differences:
\begin{itemize}
\item The free Hamiltonian is now a matrix, $\Omega_{\mathbf k} \equiv \sqrt{|\mathbf k|^2 + M^2(x)}$, and gives rise to the new commutator term  $[\Omega_{\mathbf k}, f]$. 

\item The collision term $\mathscr{C}[f,\bar f]$, evaluated explicitly in Appendix~\ref{app:coll}, has a non-trivial matrix structure involving $y$ and the density matrices for particles ($f$) and antiparticles ($\bar f$).  (Our notation $\mathscr{C}[f,\bar f]$ indicates that $\mathscr{C}$ is a functional of $f,\bar f$.)
\end{itemize}
Although Eq.~\eqref{eq:flav} is covariant under flavor rotations, it is most convenient to work in the mass basis, denoted by the subscript $m$.  Rotating to this basis, the density matrix and free Hamiltonian transform as
\be
f(\mathbf k,x) \to f_m(\mathbf k,x) = U^\dagger(x) f(\mathbf k,x) U(x) \; , \quad \Omega_{\mathbf k} \to \omega_{\mathbf k} \equiv \left( \ba{cc} \omega_{1\mathbf k} & 0 \\ 0 & \omega_{2\mathbf k} \ea \right) \; ,
\label{fltrans}
\ee
where $\omega_{i\mathbf k} \equiv \sqrt{|\mathbf k|^2 + m_i^2(x)}$.  The Boltzmann equation becomes
\be
(\partial_t + \mathbf{v} \cdot \nabla_{\mathbf x} + \mathbf{F} \cdot \nabla_{\mathbf k} ) \, f_m(\mathbf k,x) = - \left[ i \, \omega_{\mathbf k} + \Sigma^0 + \mathbf{v} \cdot \boldsymbol{\Sigma}, \, f_m(\mathbf k,x) \right] + \mathscr{C}_m[f_m, \bar f_m](\mathbf k,x) \; . \label{eq:flav2}
\ee
A similar equation governs the evolution of the antiparticle density matrix $\bar{f}_m(\mathbf k,x)$.  The quantum Boltzmann equations for $f_m$ and $\bar{f}_m$ are derived more rigorously below.  The final result  for  our Boltzmann equations is  given by Eq.~\eqref{eq:qboltz}.

\subsection{Field Theory Derivation}
\label{sec:ctp}

We derive the Boltzmann equations using non-equilibrium quantum field theory in the real time Closed Time Path (CTP) formalism~\cite{SchwingerKeldysh}.  
The arguments presented here are similar to those in our previous work, where we derived the multi-flavor Boltzmann equations for purely time-dependent scalar systems~\cite{Cirigliano:2009yt}, to which we refer the reader for greater detail.  

In the CTP formalism, the basic building blocks are the non-equilibrium  Green's functions, defined here for mass-basis fields $\phi_i$ \footnote{Although our formalism is covariant under flavor rotations, we work in the mass basis since it makes our $\epsilon$ expansion the most transparent.}:
\begin{subequations}
\begin{align}
G^t_{ij}(y,y') &= \left\langle \mathbb{T} \,  \phi_i(y)\, \phi^\dagger_j(y') \right\rangle & 
G^<_{ij}(y,y') &= \left\langle \phi^\dagger_j(y')\, \phi_i(y) \right\rangle \, , \\
G^>_{ij}(y,y') &= \left\langle \phi_i(y)\, \phi^\dagger_j(y') \right\rangle &
G^{\bar t}_{ij}(y,y') &= \left\langle \bar{\mathbb{T}}\, \phi_i(y)\, \phi^\dagger_j(y') \right\rangle \; .
\end{align}
\end{subequations}
where $\mathbb{T}\, (\bar{\mathbb{T}}$) denotes (anti)time-ordering.
These Green's functions obey Schwinger-Dyson equations, which along with the free field equations following from the Lagrangian \eq{eq:massbasis} imply the equations of motion
\begin{equation}
\begin{split}
\Bigl[\partial_y^2  + m^2(y) + 2\Sigma_\mu(y)\partial_y^\mu + \Sigma_\mu\Sigma^\mu(y) + \partial_y^\mu\Sigma_\mu(y)\Bigr] G^\gtrless (y,y') &= -i\int\! d^4 z\Bigl[ \widetilde \Pi(y,z)\widetilde G(z,y')\Bigr]^\gtrless \\
G^\gtrless (y,y')\Bigl[\overset{\leftarrow}{\partial_{y'}^2}  + m^2(y') - 2\overset{\leftarrow}{\partial_y^\mu}\Sigma_\mu(y')+ \Sigma_\mu\Sigma^\mu(y') - \partial_{y'}^\mu\Sigma_\mu(y')\Bigr]  &= -i\int\! d^4 z\Bigl[ \widetilde G(y,z)\widetilde \Pi(z,y')\Bigr]^\gtrless \; ,
\end{split}
\end{equation}
where the tildes denote matrices in CTP space,
\be
\widetilde G = \begin{pmatrix} G^t & -G^< \\ G^> & -G^{\bar t} \end{pmatrix} \quad , \quad \widetilde \Pi = \begin{pmatrix} \Pi^t & -\Pi^< \\ \Pi^> & -\Pi^{\bar t} \end{pmatrix} \; ,
\ee
and $\Pi(y,y')$ is the matrix of self-energies that appears in the Schwinger-Dyson equation.

From these equations of motion, 
one derives the so-called constraint and kinetic equations for the Wigner-transformed Green's functions
\be
G^\gtrless_{ij}(k,x) \equiv \int\! d^4 r \: e^{i k \cdot r} \, G^\gtrless_{ij}\left(x+{r}/{2}, \, x-{r}/{2} \right) \; ,
\ee
where $x\equiv (y+y')/2$ and $r\equiv y-y'$. (We will let the arguments distinguish the position space Green's function and its Wigner transform.)

The constraint equation, given by
\begin{align}
\label{constraint}
\left( 2 k^2 - \frac{\partial_{x}^2}{2} \right) G^\gtrless(k,x) \; = \; e^{-i\Diamond} \, &\biggl( \, \bigl\{ \, m^2(x)- 2i \, k \cdot \Sigma(x) + \Sigma(x)^2, \: G^\gtrless(k,x) \, \bigr\}  \; \biggr. \; \;   \\
& \; + \, i \, \bigl\{\Pi^h(k,x),G^\gtrless(k,x)\bigr\} +  i \, \bigl\{\Pi^\gtrless(k,x),G^h(k,x)\bigr\} \notag \\
& \; +  \frac{i}{2} \,\bigl[\Pi^>(k,x),G^<(k,x)\bigr] +  \frac{i}{2} \, \bigl[G^>(k,x),\Pi^<(k,x)\bigr] \, \biggr) \; , \notag 
\end{align} 
determines the shell structure of the  excitations.  The kinetic equation, given by
\begin{align}
\label{kinetic}
2 k \cdot \partial_x \, G^\gtrless(k,x) \; = \; e^{-i\Diamond} & \biggl( \, - i \, \bigl[ \, m^2(x) - 2i \, k \cdot \Sigma(x) + \Sigma(x)^2, \: G^\gtrless(k,x) \, \bigr]  \; \biggr. \\
& \; + \biggl. \, \bigl[\Pi^h(k,x),G^\gtrless(k,x)\bigr] + \bigl[\Pi^\gtrless(k,x),G^h(k,x)\bigr] \notag \\ 
& \; + \frac{1}{2}\, \bigl\{\Pi^>(k,x),G^<(k,x)\bigr\} - \frac{1}{2}\, \bigl\{\Pi^<(k,x),G^>(k,x)\}\biggr)  \;, \notag
\end{align}
governs the dynamics of the system\footnote{Eqs.~\eqref{constraint} and \eqref{kinetic} correct typos in Eqs.~(25) and (26) in Ref.~\cite{Cirigliano:2009yt}.}.  
The diamond operator $\Diamond$ is defined by
\begin{equation}
\label{diamond}
\Diamond\Bigl(A(k,x)B(k,x) \Bigr) \; = \; \frac{1}{2}\, \left(\pd{A}{x^\mu}\pd{B}{k_\mu} - \pd{A}{k_\mu}\pd{B}{x^\mu}\right)\,.
\end{equation}
The $\Pi$ functions are now the Wigner-transformed self-energies and $G^h \equiv( G^t - G^{\bar t})/2$.

Working at a fixed order in a perturbative expansion in couplings $y_{L,R}$, one can express the 
self-energies $\Pi(k,x)$ as functionals of $G(k,x)$.  
Eqs.~\eqref{constraint} and \eqref{kinetic} then describe the quantum evolution of the Wigner functions $G (k,x)$. 
However,
Eqs.~\eqref{constraint} and \eqref{kinetic}  are formidable to solve in practice.  Therefore, we simplify them by working to leading non-trivial order in $\epsilon$: $\mathcal{O}(\epsilon)$ in the kinetic equation and $\mathcal{O}(\epsilon^0)$ in the constraint equation.  This follows the spirit of the effective kinetic theory developed in Refs.~\cite{Calzetta:1986cq,Blaizot:1992gn,Arnold:2002zm,Arnold:2003zc}. Our power counting in $\epsilon$ proceeds according to the following rules:
\begin{itemize}
\item Each derivative $\partial_x$ acting on $U(x)$ or $m^2(x)$ carries one power of $\ewall$; e.g., $\Sigma^\mu$ is $\mathcal{O}(\ewall)$.
\item Each factor of the self-energy $\Pi$ carries one power of $\ecoll$.  This is equivalent to an expansion in coupling constants $y_{L,R}$.
\item Each $\Delta m^2$ carries one power of $\eosc$.  In particular, the commutator $[m^2, G^\gtrless(k,x)]$ and $G^h$ are both proportional to $\Delta m^2$ and are $\mathcal{O}(\eosc)$~\cite{Cirigliano:2009yt}.
\end{itemize}
According to these rules, all terms on the right side of Eq.~\eqref{kinetic} are at least linear in $\epsilon$.  Therefore, the kinetic equation implies one more rule:
\begin{itemize}
\item Each derivative $\partial_x$ acting on $G^\gtrless(k,x)$ carries one power of $\epsilon$.
\end{itemize}
The dimensionality of these quantities (e.g., $\Delta m^2$, $
\partial_x U$, etc.) is compensated by powers of $\omega_{i\mathbf k}$ or $|\mathbf k|$, taken to be $\mathcal{O}(T)$, to 
form dimensionless ratios $\epsilon$.  Our $\epsilon$ expansion, therefore, breaks down for infrared modes $|\mathbf k| \ll 
T$.  We neglect this complication since the density of states for these modes, $k^2 f(k)$, is suppressed compared to typical thermal modes $| \mathbf k| \sim T$.

Using the above rules, the constraint equation at $\mathcal{O}(\epsilon^0)$ becomes trivial
\footnote{In general there exists a more complicated shell structure that deserves mention~\cite{Herranen:2008hi}.  For free fields (setting $\Sigma,\Pi \to 0$), it is straight-forward to solve the constraint equation to all orders in $\eosc$.  In the rest frame of the wall, for a given component $G^\gtrless_{ij}(k,x)$, there exist not two but four shells, which can be expressed as (using over-bar for quantities in the wall rest-frame)
\be
\bar{k}_z = \pm (\bar{k}_{zi} + \bar{k}_{zj})/2 \; , \quad  \bar{k}_z = \pm (\bar{k}_{zi} -\bar{ k}_{zj})/2 \; , \notag
\ee
where $\bar{k}_{zi}\equiv\sqrt{\bar{k}_0^2 - k_x^2 - k_y^2 - m_i^2}$.  The $\bar{k}_z\!=\!\pm (\bar{k}_{zi} + \bar{k}_{zj})/2$ shells  
describe coherence of states moving in the same direction  (for $i \neq j$ these are different eigenstates) 
and reduce to Eq.~\eqref{freeconst} for $\eosc\!=\!0$. 
The other shells $\bar{k}_z = \pm (\bar{k}_{zi}\! -\! \bar{k}_{zj})/2$, named ``non-local coherence shells,'' correspond to coherence between states of opposite momentum (and for $i \neq j$ different mass eigenstates), and are interpreted in terms of quantum mechanical reflection~\cite{Herranen:2008hi}.  By performing a mode expansion of $G^\gtrless$ in terms of free-field creation and annihilation operators, it is possible to show that the coherence shells arise from non-zero expectation values $\langle a_i b_j \rangle$ and $\langle a^\dagger_i b^\dagger_j \rangle$; such an effect is also known as {\it zitterbewegung} \cite{zbw}.  In the thick wall regime $(L_{w} \gg L_{\textrm{int}})$ of EWBG, we expect on physical grounds that occupation numbers associated with reflection should be suppressed and we neglect these shells in our analysis.  However, their importance in thin wall regime $(L_{w} \lesssim L_{\textrm{int}})$ has been emphasized in Refs.~\cite{Herranen:2008hi}.
}:
\be
\left(k^2 - \bar{m}^2(x) \right) G^\gtrless(k,x) = 0 \; ,
\ee
where $\bar{m}^2 \equiv ( m_1^2 + m_2^2)/2$.  Therefore, $G^\gtrless(k,x)$ must vanish unless 
\be
k^0 = \pm \, \bar{\omega}_{\mathbf k}(x) = \pm \sqrt{ |\mathbf k|^2 + \bar{m}^2(x) } \; . \label{freeconst}
\ee
The two shells correspond to particle ($k^0\! >\!0$) and antiparticle ($k^0\! <\!0$) modes, and the two-point functions 
can be expressed in terms of particle ($f_m$) and antiparticle ($\bar{f}_m$) mass basis density matrices as follows:
\begin{equation}
\label{eq:treesolutionv1}
\begin{split}
G^{>}(k,x)  &= 2\pi\delta(k^2 - \bar{m}^2) \, \left[\, \theta(k^0)(I + f_m(\mathbf k, x)) + \theta(-k^0) \bar {f}_m(-\mathbf k,x) \, \right] ~,  \\
G^{<}(k,x)  &= 2\pi\delta(k^2 - \bar{m}^2) \, \left[\,\theta(k^0) f_m (\mathbf k,x) + \theta(-k^0)(I + \bar f_m (-\mathbf k,x)\, \right] \, ,
\end{split}
\end{equation}
where $I$ is the $2 \times 2$ identity matrix.

The Boltzmann equation is obtained from the kinetic equation~\eqref{kinetic}.  
Working to $\mathcal{O}(\epsilon)$, we have
\begin{align}
\label{kinetic2}
&2 k \cdot \partial_x \, G^\gtrless(k,x) \; = \;   - \bigl[ \, i \, m^2(x) + 2 \, k \cdot \Sigma(x) - \Pi^h(k,x), \: G^\gtrless(k,x) \, \bigr]  \\
& \quad + \frac{1}{2} \left\{ \partial_x^\mu m^2, \, \partial_{k^\mu} G^<(k,x) \right\} + \frac{1}{2}\, \bigl\{\Pi^>(k,x),G^<(k,x)\bigr\} - \frac{1}{2}\, \bigl\{\Pi^<(k,x),G^>(k,x)\} \;. \notag
\end{align}
By taking the positive (negative) frequency integrals of $G^\gtrless(k,x)$, we can project out the particle (antiparticle) density matrices:
\be
f_m(\mathbf k, x) = \int^\infty_0 \! \frac{dk^0}{2\pi} \: 2k^0 \, G^<(k,x)  \; , \quad 
\bar{f}_m(-\mathbf k, x) = \int_{-\infty}^0 \! \frac{dk^0}{2\pi} \: (-2k^0) \, G^>(k,x) \; .
\ee
Taking the positive frequency integral of Eq.~\eqref{kinetic2}, we arrive at the Boltzmann equations for $f_m$.  Here, a useful relation is  
\be
\int^\infty_0 \! \frac{dk^0}{2\pi} \: G^<(k,x) = \int^\infty_0 \! \frac{dk^0}{2\pi} \: \left( \frac{2k^0}{2\bar{\omega}_{\mathbf k}} \right) \, G^<(k,x) \;  + \; \mathcal{O}(\epsilon)  = \frac{f(\mathbf k,x)}{2\bar \omega_{\mathbf k}} + \mathcal{O}(\epsilon)\; ,
\ee
using Eq.~\eqref{freeconst}, according to which the factor $(k^0/\bar\omega_{\mathbf k})$ is equal to unity (restricted to $k^0 \! > \! 0$), modulo $\mathcal{O}(\epsilon)$ corrections. These corrections can be neglected since we are working to linear order in $\epsilon$ and every term in Eq.~\eqref{kinetic2} is already $\mathcal{O}(\epsilon)$.

We now evaluate the various terms in the Boltzmann equation.
The left side of Eq.~\eqref{kinetic2} is
\be
\int^\infty_0 \frac{dk^0}{2\pi} \,  2k\cdot \partial_x G^<(k,x) =  
\left(\partial_{t} + \mathbf{v} \cdot \nabla_{\mathbf x} \right) f_m(\mathbf k,x) + \mathcal{O}(\epsilon^2)\; ,
\ee
with velocity $\mathbf{v}\! = \! \mathbf k/\bar{\omega}_{\mathbf k}$.
The oscillation term is
\be
\int^\infty_0 \! \frac{dk^0}{2\pi}  \, \left[ m^2(x), \, G^<(k,x) \right] = \left[ \omega_{\mathbf k}, f_m(\mathbf k,x) \right] + \mathcal{O}(\epsilon^2) \; ,
\ee
using the fact that $(m_1^2-m_2^2)/(2\bar{\omega}_{\mathbf k}) = (\omega_{1\mathbf k} - \omega_{2\mathbf k})$.  
 The CP-violating source term is
\be
\int^\infty_0 \frac{dk^0}{2\pi} \, \left[2k\cdot \Sigma(x),  G^<(k,x) \right] = \left[ \Sigma^0(x) + \mathbf{v} \cdot \boldsymbol{\Sigma}(x), \, f_m(\mathbf k,x) \right] + \mathcal{O}(\epsilon^2)\; .
\ee
and the force term is
\be
\int^\infty_0 \frac{dk^0}{2\pi} \, \frac{1}{2} \left\{ \partial_x^\mu m^2, \, \partial_{k^\mu} G^<(k,x) \right\} = - \, \mathbf{F} \cdot \nabla_{\mathbf k} f_m(\mathbf k,x) + \mathcal{O}(\epsilon^2) \; , \label{force}
\ee
with force $\mathbf F = - \nabla_{\mathbf x} \bar{\omega}_{\mathbf k}$. (The $\partial_{k^0}$ contribution to Eq.~\eqref{force} is a total derivative and vanishes at the boundaries.)

The remaining terms in Eq.~\eqref{kinetic2}, arising from the self-energy $\Pi$, give two important contributions (see Appendix~\ref{app:coll}).  First, the $[\Pi^h, G^\gtrless]$ term yields a medium-dependent, forward-scattering correction to the mass matrix.  For the interaction given in Eq.~\eqref{eq:lint}, assuming the $A$ bosons are in thermal equilibrium, this correction gives the thermal mass shift $m_{L,R}^2 \to m_{L,R}^2 + y_{L,R} \, T^2/24$.  This shift can be incorporated directly into $m(x)$ and $\Sigma^\mu(x)$.  The remaining collision term
\be
\mathscr{C}_m[f_m,\bar f_m] \; \equiv \; \int^\infty_0 \! \frac{dk^0}{2\pi} \, \frac{1}{2} \, \left(  \bigl\{\Pi^>(k,x),G^<(k,x)\bigr\} - \bigl\{\Pi^<(k,x),G^>(k,x) \} \, \right) \label{colldef}
\ee
corresponds to scattering $(\phi A \leftrightarrow \phi A)$ and annihilation $(\phi \phi^\dagger \leftrightarrow A A$) processes in the plasma.

To summarize, the quantum Boltzmann equations are\footnote{We follow a convention where the antiparticle density matrix obeys the same flavor transformation rule $\bar f \to U^\dagger \bar f_m U$ as $f$ in Eq.~\eqref{fltrans}.  If we evaluated the fields $\phi, \phi^\dagger$ in terms of creation/annihilation operators, the density matrices would be $(f_m)_{ij} \sim \langle a^\dagger_j a_i \rangle$ and $(\bar f_m)_{ij} \sim \langle b^\dagger_i b_j \rangle$. The swapping of $i,j$ between $f_m$ and $\bar f_m$ is the reason for the sign flip between the terms $[\omega_{\mathbf k}, f_m]$ and $[\omega_{\mathbf k}, \bar f_m]$ in Eq.~\eqref{eq:qboltz}.  
If we chose the alternate transformation convention $\bar f \to U^\top \bar f_m U^*$, we would have $(\bar f_m)_{ij} \sim \langle b^\dagger_j b_i \rangle$, no sign-flipped $[\omega_{\mathbf k}, \bar f_m]$ term, and $\Sigma^\mu$ replaced by $\Sigma^{\mu *}$.\label{foot1}}
\begin{subequations}
\label{eq:qboltz}
\bea
(u \cdot \partial_x + \mathbf{F} \cdot \nabla_{\mathbf k} ) \, f_m(\mathbf k,x) &=& - \,\left[ i \, \omega_{\mathbf k} + u \cdot \Sigma, \, f_m(\mathbf k,x) \right] + \mathscr{C}_m[f_m, \bar f_m](\mathbf k,x) \\
(u \cdot \partial_x + \mathbf{F} \cdot \nabla_{\mathbf k} ) \, \bar f_m(\mathbf k,x) &=& + \left[ i \, \omega_{\mathbf k} - u \cdot  \Sigma, \, \bar f_m(\mathbf k,x) \right] + \mathscr{C}_m[\bar f_m, f_m](\mathbf k,x)
\eea
\end{subequations}
for the (anti)particle density matrix $f_m$ ($\bar f_m$), with $u^\mu\equiv (1,\mathbf v)$ and $\partial_x^\mu \equiv (\partial_t,\nabla_{\mathbf x})$.  

These equations are identical in structure to the usual single-flavor Boltzmann equations, with two additional ingredients.  First, the term $[i\omega_{\mathbf k} , f_m]$ gives rise to $(\Phi_L,\Phi_R)$ flavor oscillations.  Second, the $[u \cdot \Sigma,f_m]$ term is the CP-violating source, due to spacetime-dependent mixing.  This term is a ``source'' because it does not vanish when $f_m,\bar f_m$ are in equilibrium; furthermore, it violates C and CP symmetries, as we show below.  Lastly, it is straightforward to show that Eqs.~\eqref{eq:qboltz} are consistent with the continuity equation for the total $\Phi_L + \Phi_R$ charge, $\textrm{Tr}\int d^3k/(2\pi)^3 (u \cdot \partial_x) (f_m-\bar f_m) = 0$.

\subsection{C and CP violation}

It is insightful to consider how C and CP violation are manifested in the Boltzmann equations \eqref{eq:qboltz}.  (See Ref.~\cite{Cirigliano:2009yt} for a complementary discussion at the Lagrangian level.)  Under C, the density matrices transform as \footnote{The reason for the transpose is our convention for $\bar f_m$ given in footnote~\ref{foot1}.  In the alternate convention, one would have $f_m \; \xrightarrow{\;\; \textrm{C} \; \; } \; \eta  \bar f_m  \eta^\dagger$.
}
\be
f_m(\mathbf k,x) \; \xrightarrow{\;\; \textrm{C} \; \; } \; e^{i\eta} \, \bar f_m^{\, \top}(\mathbf k,x)  \, e^{-i\eta} \; , \label{Ctrans}
\ee
where $\eta \equiv \textrm{diag}(\eta_1, \eta_2)$ are arbitrary phases.  The Boltzmann equations are C-symmetric if
\be
e^{i\eta} \, \Sigma_\mu^{\,\top} \, e^{-i\eta} = - \; \Sigma_\mu \; 
\ee
for some choice of $\eta$.  Therefore, from Eq.~\eqref{eq:sigma}, C violation requires $\sin \theta \ne 0$ and $\partial_\mu \alpha \not = 0$.

Under P, the density matrices transform as
\be
f_m(\mathbf k,x) \; \xrightarrow{\;\; \textrm{P} \; \; } \; e^{i\bar\eta} \, f_m(-\mathbf k,x^\prime) \, e^{-i\bar\eta} \; , \label{Ptrans}
\ee
where $\bar\eta \equiv \textrm{diag}(\bar\eta_1,\bar\eta_2)$ are again arbitrary phases, and $x^{\prime\mu} \equiv (t, - \mathbf x)$ is the P-inverted coordinate.  In general, inhomogeneous background fields break spatial symmetries, such as P.  However, a spherical bubble centered at $r=0$ is invariant under P: therefore, we have
\be
m^2(x) = m^2(x^\prime) \, , \;\; U(x) =  U(x^\prime) \, , \;\; \mathbf{F}(x) = - \mathbf{F}(x^\prime) \, , \;\; \Sigma^\mu(x) = \left(\Sigma^0(x^\prime), - \boldsymbol{\Sigma}(x^\prime) \right) \, ,
\ee
using the fact that $\nabla_{\mathbf x} = - \nabla_{\mathbf x^\prime}$.  The Boltzmann equation for $f_m$ transforms under P into
\begin{align}
(\partial_t  - \mathbf{v} \cdot \nabla_{\mathbf x^\prime} - & \, \mathbf{F} \cdot \nabla_{\mathbf k} ) \, f_m(-\mathbf k,x^\prime)  \notag \\
& = - \,\left[ i \, \omega_{\mathbf k}(x^\prime) + \Sigma^0(x^\prime) - \mathbf{v} \cdot \boldsymbol{\Sigma}(x^\prime), \, f_m(-\mathbf k,x^\prime) \right] + \mathscr{C}_m[f_m, \bar f_m](-\mathbf k,x^\prime) \; ,
\end{align}
taking $\bar{\eta}_{1,2}\!=\! 0$.  Therefore, setting $\mathbf k \to - \mathbf k$ and relabeling $x^\prime \to x$, we find that Eqs.~\eqref{eq:qboltz} are invariant under P. 

In summary, the Sakharov conditions of C and CP violation are realized if the bubble wall induces flavor mixing ($\sin\theta \ne 0$) and a spacetime-dependent phase ($\partial_\mu \alpha \ne 0$) in the two-scalar system.  For a spherical bubble, C and CP violation are equivalent, since P is conserved.  An aspherical bubble will in general violate P, and therefore CP, but clearly this is insufficient by itself for EWBG if C is conserved.\footnote{Later we will take the planar limit of the bubble wall as given by \eq{eq:kink}, which apparently violates P. However, \eq{eq:kink} really only describes one ``side'' of the bubble in the planar limit, with the other side infinitely far away, hiding its true parity invariance. Under P, \eq{eq:kink}  remains unchanged.}


\section{Solution to the flavored Boltzmann equations}
\label{Sect:solution}

In this section, we solve the flavored Boltzmann equations numerically, 
organizing our discussion as follows. 
In Section  \ref{Sect:solutionA}   
we apply the formalism of Sec.~\ref{Sect:QBE} to our EWBG toy model 
of mixing scalars $\Phi$ in a bubble wall geometry, and we describe what are the quantities of interest for EWBG.  
In  Section \ref{Sect:solutionB}   we show that EWBG shares a physical analogy with spin precession in a varying magnetic field.
In Section  \ref{Sect:solutionC} we describe our numerical methods for solving the Boltzmann equations.  We
present our numerical results  and discuss the implications for EWBG
in Section  \ref{sec:results}.
Our main conclusions are:
\begin{itemize}
\item We demonstrate what are the key charge transport dynamics of EWBG: CP-violating flavor mixing and coherent oscillations in a spacetime-dependent background, collisional damping that destroys this coherence, and diffusion of charge into the unbroken phase.  Our solutions are exact, without any ansatz for the functional form of the $\Phi$ density matrices $f,\bar{f}$.
\item  We find a resonant enhancement of CP-violating charge generation for $m_L \sim m_R$. 
Similar resonances were discussed previously in Refs.~\cite{Carena:1997gx,Riotto:1998zb,Carena:2000id,Carena:2002ss,Balazs:2004ae,Lee:2004we,Cirigliano:2006wh}, but our  present work establishes its true origin on more theoretically sound footing.
The width and height of this resonance are  controlled by 
the ratio  $L_{\textrm{osc}} / L_{{w}}$ for typical thermal modes.
\item CP-violating charge {\it does} diffuse into the unbroken phase.  This can provide a potentially large enhancement of charge production compared to previous treatments of EWBG flavor oscillations in Ref.~\cite{konstandin}, which found that diffusion of the oscillating species was quenched.  We identify the reason for this discrepancy in Sec.~\ref{Sect:comparison}, finding it is due to the breakdown of the power counting expansion followed in Ref.~\cite{konstandin}. 
\end{itemize}

To clarify our strategy for power counting solutions to the Boltzmann equations, we distinguish between power counting terms in the equations themselves, and   explicitly expanding the solutions for $G^\gtrless(k,x)$ perturbatively in $\epsilon$. We perform the former but  not the latter. For the equations of motion at a given order in $\epsilon$, we will solve for $G(k,x)$ exactly as a function of $\epsilon$. We define the $\mathcal{O}(\epsilon^n)$  solution for $G(k,x)$ as the exact solution of the $\mathcal{O}(\epsilon^n)$ equation of motion. To be precise, we will obtain the exact solutions of the constraint equation at $\mathcal{O}(\epsilon^0)$ and the kinetic equation to $\mathcal{O}(\epsilon)$, an appropriate strategy to solve for the leading nontrivial deviations of the distribution functions away from equilibrium. This method avoids making any \emph{a priori} ansatz about the functional form or power counting of $G(k,x)$.\footnote{Furthermore, it is only by this method that one obtains solutions that equilibrate properly at late time (or far from the bubble wall). Otherwise we run into the problem of ``secular terms'' that grow large and spoil equilibration at late time \cite{Berges:2004yj}. Our procedure will not encounter any such problematic terms.}

\subsection{EWBG setup}
\label{Sect:solutionA}

For EWBG in a late time regime (compared to the bubble nucleation time), significant simplifications arise in solving the quantum Boltzmann equations~\eqref{eq:qboltz}.  First, we neglect the wall curvature, treating the spherical bubble as a planar wall, where $z<0$ corresponds to the broken phase inside the bubble, as given by Eq.~\eqref{eq:kink}.   Second, we look for steady state solutions in the rest frame of the moving wall (with $v_w \ll 1$) for the $\Phi$ particle and antiparticle density matrices $f(\mathbf k,z)$ and $\bar{f}(\mathbf k,z)$, assumed to be a function only of $z = r - v_w t$, the coordinate normal to the wall.  Additionally, the force term $\mathbf F$ vanishes in our model since $m_1^2+m_2^2$ is constant.  Thus, Eq.~\eqref{eq:qboltz} becomes\footnote{Henceforth, we work in the mass basis and drop the $m$ subscripts whenever it does not lead to ambiguitues.}
\begin{subequations}
\bea
v_{\textrm{rel}} \, \partial_z f(\mathbf k,z) &=& - \bigl[ i \, \omega_{\mathbf k} + v_{\textrm{rel}} \, \Sigma, \, f(\mathbf k,z) \bigr] + \mathscr{C}[f,\bar f](\mathbf k,z) \; \\
v_{\textrm{rel}} \, \partial_z \bar{f}(\mathbf k,z) &=& \ \  \bigl[ i \, \omega_{\mathbf k} - v_{\textrm{rel}} \, \Sigma, \, \bar f(\mathbf k,z)\bigr] +  \mathscr{C}[\bar f,f](\mathbf k,z) \; ,
\eea
\label{eq:QBE1}
\end{subequations}
with 
\be
v_{\textrm{rel}}(\mathbf k) \equiv \frac{\mathbf k \cdot \hat{\mathbf n}}{\bar{\omega}_{\mathbf k}} - v_w \; ,  \quad
\omega_{\mathbf k}(z) \equiv \left( \ba{cc} \omega_{1{\mathbf k}}(z) & 0 \\ 0 & \omega_{2\mathbf k}(z) \ea \right) \; , \quad
\Sigma(z) \equiv U^\dagger(z) \partial_z U(z) \; ,
\ee
where $v_{\textrm{rel}}$ is the velocity with respect to the wall and $\hat{\mathbf{n}}$ is the unit vector normal to the wall. From now on, the collision term $\mathscr{C}$ includes a factor of $g_*$ to model the true number of degrees of freedom in the electroweak plasma and is given by \eq{totalcollision}.
Given a set of input model parameters, 
we want to solve Eq.~\eqref{eq:QBE1} subject to the boundary condition that far from the wall the solutions reach equilibrium: 
\be \label{eq:boundary}
\lim_{z \to \pm \infty} f(\mathbf k,z) , \, \bar{f}(\mathbf k,z) = \lim_{z \to \pm \infty} f^{\textrm{eq}}(\mathbf k,z) \, , \quad
f^{\textrm{eq}}(\mathbf k,z) \equiv  \left( \ba{cc} n_B(\omega_{1\mathbf k}(z)) & 0 \\ 0 & n_B(\omega_{2\mathbf k}(z)) \ea \right) \; .
\ee
Since Eq.~\eqref{eq:QBE1} has azimuthal symmetry with respect to $\hat{\mathbf{n}}$,  
the density matrices 
 $f,\bar{f}$ depend only on the momentum variables $k\equiv|\mathbf k|$ and $\cos\vartheta_k \equiv \hat{\mathbf k} \cdot  \hat{\mathbf n}$, and are  independent of the azimuthal $\mathbf k$ angle.

After solving Eq.~\eqref{eq:QBE1}, we compute total charge asymmetries that are directly relevant for EWBG. The charge current matrix (in the mass basis) can be defined equivalently in terms of normal-ordered fields, Green's functions, or density matrices:
\bea 
j_m^\mu(x)_{ij} &\equiv& i  \langle \, : \phi^\dagger_j(x) \,\smash{\overset{\leftrightarrow}{\partial}}^\mu_x \phi_i(x) :\, \rangle = 
 \int \frac{d^4 k}{(2 \pi)^4}  \, k^\mu \, 
 \left(  G_{ij}^< (k,x)    +  G_{ij}^> (k,x) \right)    
 \notag \\
&=& \int \! \frac{d^3 k}{(2\pi)^3} \, \frac{k^\mu}{\bar{\omega}_{\mathbf k}} \,  \left(f_m(\mathbf k,x) - \bar{f}_m(\mathbf k,x) \right)_{ij} \; .
\label{eq:currentdef}
\eea
The analogous flavor-basis charge current is $j^\mu_{\textrm{fl}} = U \, j_m^\mu \, U^\dagger$, and this coincides with $j_m^\mu$ in the unbroken phase ($z \! > \! 0$).  The quantities
\be 
n_1(z) \equiv (j^0_m)_{11} \, , \; n_2(z) \equiv (j^0_m)_{22} \, , \; n_L(z) \equiv (j^0_{\textrm{fl}})_{LL} \, , \; n_R(z) \equiv (j^0_{\textrm{fl}})_{RR}
\ee
are the spacetime-dependent total charge densities of $\phi_1$, $\phi_2$, $\Phi_L$, and $\Phi_R$ states, respectively.  In the unbroken phase, $n_{1,2} = n_{L,R}$.  In addition, we define the integrals
\be
{I}^{CP}_{L,R} = \int^\infty_0 dz \, n_{L,R}(z)
\label{eq:ILR}
\ee
as the total charge in the unbroken phase. 
This is a useful global measure of CP violation and a suitable proxy for the baryon asymmetry itself.
For example, in squark-driven EWBG scenarios~\cite{Blum:2010by}, where $\Phi = (\widetilde t_L, \widetilde t_R)$, ${I}^{CP}_{L}$ and ${I}^{CP}_{R}$ will be converted into left-handed quark charge through fast gaugino- and Higgsino-mediated processes, respectively, thereby sourcing baryon generation through electroweak sphalerons (which are only active in the unbroken phase).  Therefore, baryon number will be directly proportional to ${I}^{CP}_{L,R}$.  (A more precise statement requires a ``realistic'' model such as~\cite{Blum:2010by}, beyond the scope of this work.)

\subsection{Magnetic analogy}
\label{Sect:solutionB}

At this point, it insightful to introduce an analogy with spin precession in a varying magnetic field \cite{Stodolsky:1986dx}.  
The $2 \times 2$ Hermitian density matrices $f,\bar{f}$ can be expressed by Bloch 
decomposition as four-vectors $p= (p_0, \mathbf p)$ and $\bar p= (\bar p_0, \bar{\mathbf p})$:
\be
\label{Bloch}
f(\mathbf k,z) = I \, p_0(\mathbf k,z) + \boldsymbol{\sigma} \cdot \mathbf{p}(\mathbf k,z) \; , \qquad
\bar{f}(\mathbf k,z) = I \, \bar{p}_0(\mathbf k,z) + \boldsymbol{\sigma} \cdot \bar{\mathbf{p}}(\mathbf k,z) \; ,
\ee
where $\boldsymbol{\sigma}\equiv (\sigma_1, \sigma_2, \sigma_3)$ are Pauli matrices and $I$ is the identity.  $p_0(\mathbf k,z)$ represents the total occupation number of all $\phi_1 + \phi_2$ particles, while the ``polarization'' vector $\mathbf{p}(\mathbf k,z)$ describes the density matrix for the internal flavor degrees of freedom, for a given momentum $\mathbf k$ (and similarly with $\bar{p}(\mathbf k,z)$ for anti-particles)\footnote{
We denote unit vectors in the internal flavor space by $(\hat{\mathbf x}, \hat{\mathbf y}, \hat{\mathbf 
z})$, not to be confused the spatial coordinate $z$ with respect to the wall.
}.
One can also decompose the collision term as
\be
\frac{1}{v_{\rm rel}} \, \mathscr{C}[f,\bar f](\mathbf k,z)  
= - I \, D_0 [p, \bar{p}](\mathbf k,z) - \boldsymbol{\sigma} \cdot \mathbf{D}[p, \bar{p}](\mathbf k,z)  \; 
\ee
to define the damping vector $D = (D_0, \mathbf{D})$.

In the polarization vector language, Eq.~\eqref{eq:QBE1} becomes
\begin{subequations}
\label{eq:precession1}
\bea
\partial_z \mathbf{p} (\mathbf k,z) &= & ( \mathbf{B}_0   +  \mathbf{B}_\Sigma ) \times\mathbf{p} (\mathbf k,z)  
-   \mathbf{D}[p, \bar{p}](\mathbf k,z) 
 \\
\partial_z \bar{\mathbf{p}} (\mathbf k,z) &= & - ( \mathbf{B}_0   -  \mathbf{B}_\Sigma ) \times \bar{\mathbf{p}} (\mathbf k,z)  
-   \mathbf{D}[\bar p, {p}](\mathbf k,z) 
\eea
\end{subequations}
with  effective  magnetic field given by 
\begin{subequations}
\bea
\mathbf{B}_0(\mathbf k,z) &=&   \Big( 0, 0, \frac{\omega_{1\mathbf k}(z) - \omega_{2\mathbf k}(z)}{v_{\rm rel} (\mathbf k)} \Big) \\
\mathbf{B}_\Sigma  (z)  &=& \left( 2 \sin \alpha \ {\theta'} + \sin 2 \theta \, \cos \alpha \, {\alpha'}, 
\ - 2 \cos \alpha \, {\theta'}  + \sin 2\theta \, \sin \alpha \, {\alpha'},  \ 2 \sin^2 \theta \, {\alpha'} 
\right)\; ,
\eea
\end{subequations}
where the primes $'$ denote derivatives $d/dz$.
The equations of motion are strongly suggestive of spin precession in a magnetic field.  Here, flavor polarizations $\mathbf{p}, \bar{\mathbf p}$ play the role of spin and precess around an effective magnetic field $( \mathbf{B}_0 \pm \mathbf{B}_\Sigma )$ in flavor space.  This describes coherent flavor oscillations.  The collision term $\mathbf D$ destroys coherent oscillations by damping the precession.  The total particle/antiparticle occupation numbers obey Boltzmann-type equations with collisions 
\be
\partial_z {p}_0 (\mathbf k,z) =  - D_0 [p, \bar{p}](\mathbf k,z) \; , \qquad
\partial_z \bar{\mathbf p}_0 (k,z) =  - D_0 [\bar{p},p](\mathbf k,z) \; .
\ee

\begin{figure}[t]
\begin{center}
\vspace{0.25cm}
\mbox{\hspace*{-.75cm}\epsfig{file=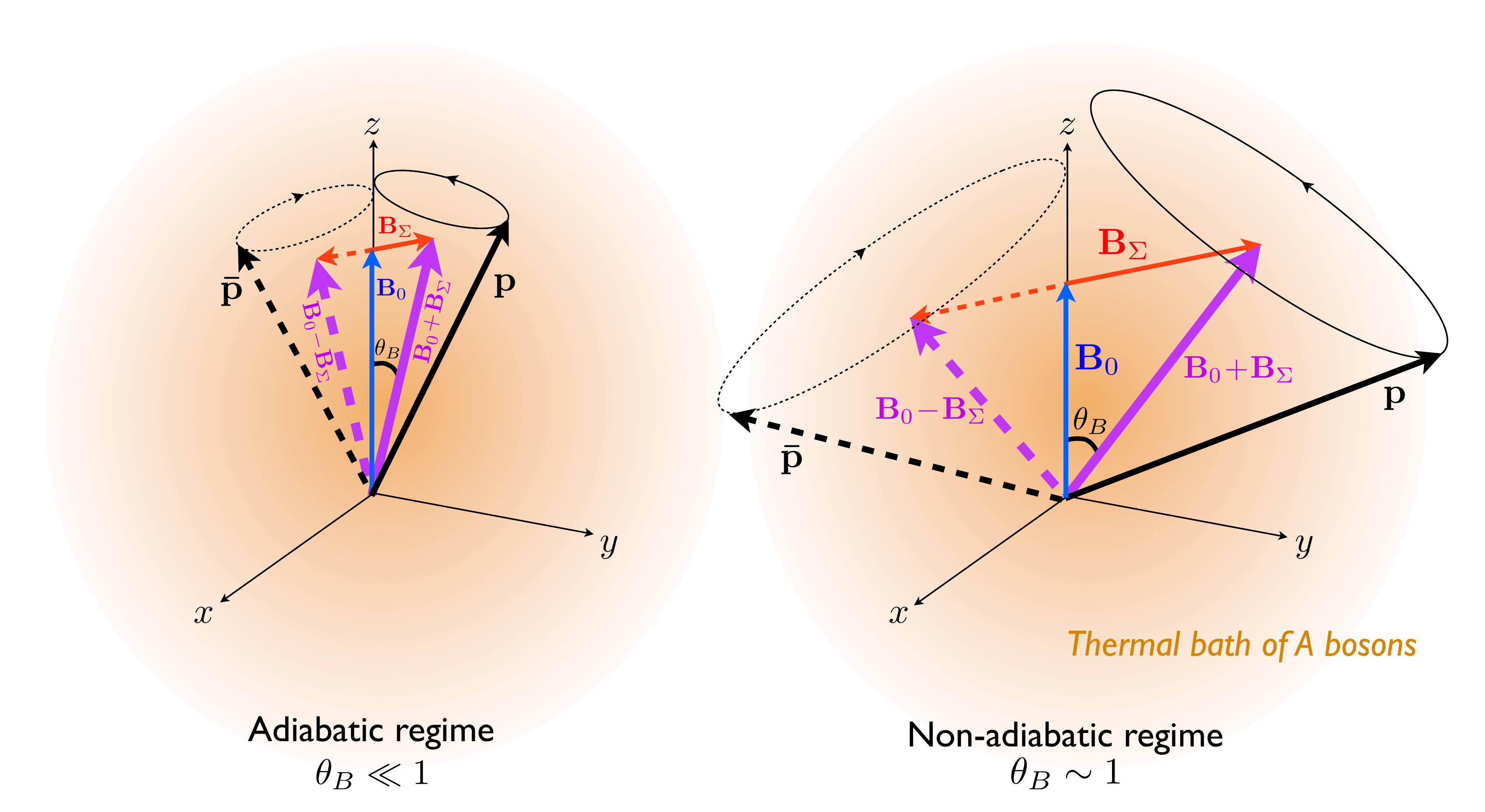,width=17.75cm}}
\end{center}
\vspace{0cm}
\caption{\it\small 
Precession of flavor polarization vectors about effective magnetic fields in the mass basis. In equilibrium, $\vect{p}= \bar{\vect p} = (0,0,n_B(\omega_1)-n_B(\omega_2))/2$ and the magnetic field $\vect{B}_0=(0,0,(\omega_1-\omega_2)/v_{\text{rel}})$ both point along the $\hat{\vect z}$ direction, and there is no precession. When the bubble wall turns on, it induces an additional magnetic field $\vect{B}_\Sigma$, causing $\vect{p},\vect{\bar p}$ to precess around $\vect{B}_0\pm \vect{B}_\Sigma$, which corresponds to flavor oscillations. If  $\alpha' \neq 0$ so that there is CP-violation,  $\vect{p},\vect{\bar p}$ develop different $\hat{\vect z}$ components, generating CP-asymmetric  diagonal densities. In the adiabatic regime, $\vect{B}_\Sigma\ll \vect{B}_0$, the angle of precession is small, the system remains near equilibrium, and large CP asymmetries are not generated. In the non-adiabatic regime, $\vect{B}_\Sigma\gtrsim \vect{B}_0$, the precession angle is large, and large deviations from equilibrium and large CP asymmetries can arise. The precession is damped by collisions with $A$ bosons in the thermal bath, leading the system back to equilibrium at late time.
}
\label{fig:precession}
\end{figure}

This analogy provides an intuitive framework to understand the 
behavior of our numerical results. We illustrate this picture in Fig.~\ref{fig:precession}.
The qualitative features of the solutions are controlled by two ratios of scales: 
\begin{itemize}
\item \textbf{Oscillation vs. wall length:} the ratio $L_{\rm osc}/L_{w}$, with  oscillation length $L_{\rm osc} = 2 \pi \,  v_{\rm rel}/(\omega_{1\mathbf k} - \omega_{2\mathbf k})$ 
and wall thickness $L_{w}$ (which determines  the $z$-dependence  of the effective magnetic field), controls how quickly the magnetic field varies on the intrinsic time scale of the system. 
In the {\it adiabatic} regime ($L_{\rm osc}/L_{w} \ll 1$),  
the polarization vector tracks the magnetic field with small amplitude precession.  In the {\it non-adiabatic} regime ($L_{\rm osc}/L_{w} \geq 1$), 
the polarization vector precesses with large amplitude. 
\item \textbf{Collisional mean free path vs. wall length:} the ratio $L_{\rm coll} / L_{w}$, with typical collision mean free path $L_{\rm coll}$, controls how fast the CP asymmetry is damped away and equilibrium is restored away from the wall.  In the overdamped regime, ($L_{\rm coll}/L_{w} \ll 1$) precession is efficiently damped away, allowing no CP asymmetry to develop.  In our analysis, we consider $L_{\rm coll}/L_{w} \geq 1$; interactions do not 
affect sizably the sourcing of CP asymmetries, but re-establish equilibrium away from the wall ($|z| \gg L_{w}$). 
\end{itemize}
These above concepts will serve as organizing principles for the discussion in the following sections. 

Lastly, we note
\be
n_{1,2}(z) = \frac{1}{2} \int \frac{d^3 k}{(2\pi)^3} \left( p_0(\mathbf k,z) \pm p_z(\mathbf k,z) - \bar{p}_0(\mathbf k,z) \mp \bar p_z(\mathbf k,z) \right) \; .
\ee
That is, the charge densities are determined by the differences $p_z\!-\!\bar{p}_z$ and (to a lesser extent, as it turns out) $p_0\!-\!\bar{p}_0$. In equilibrium, $\mathbf{p}$ and $\bar{\mathbf{p}}$ are aligned with $\mathbf{B}_0$, and $p_0 = \bar{p}_0 = (n_B(\omega_{1\mathbf k})+n_B(\omega_{2\mathbf k}))/2$ and $p_z = \bar{p}_z = (n_B(\omega_{1\mathbf k})-n_B(\omega_{2\mathbf k}))/2$.

\subsection{Numerical approach} 
\label{Sect:solutionC}

\begin{table}[t!]
\begin{center}
\begin{tabular}{|ccccccccc|}
\hline 
 &  & & &   &  &  & & \\
$\ L_{w} = \frac{20}{T}   \ $ &  
$v_w = \frac{1}{20} \ \ $ &     
$v_0 = T^2    \ \ $       &  
$\alpha_0= \frac{\pi}{2} \  \ $  &   
$m_L = 2.2\, T \ \ $ & 
$m_R= 2 \,  T \ \ $  &   $y_L = 1 \ \ $  &  $y_R = 0.75 \ \ $  & $g_* = 200\ $   \\
 &  & & &   &   &  & & \\
\hline
\end{tabular}
\end{center}
\caption{Parameters  that define the ``baseline model"  used to illustrate the main features 
of the numerical solution.   All dimensionful parameters are expressed in 
units of the temperature $T$ or its inverse.  In addition, we assume thermal masses  $(m_A/T)^2 =  (y_L + y_R)/12$
for the  scalar field $A$.}
\label{tab:baseline}
\end{table}

The Boltzmann equations, equivalently described by Eq.~\eqref{eq:QBE1} and Eq.~\eqref{eq:precession1}, are a system of 8 coupled integro-differential equations, due to the collision term coupling together modes of different momenta.  In order to make this problem tractable, we discretize $k \equiv |\mathbf k|$ and $\cos\vartheta_k$ into $N_k$ and $N_\vartheta$ bins, within the ranges
\be
0 < k < k_{\rm max} \; , \qquad -1 < \cos\vartheta_k < 1 \; ,
\ee
evaluating the discretized $(k,\cos\vartheta_k)$ at the cental value of each bin.  After binning, we have a system of $8 \times N_k \times N_\vartheta$ coupled first order ordinary differential equations with boundary conditions. 
We solve this system of equations using the ``relaxation method''~\cite{Press:1992zz}.  

We impose the boundary conditions~\eqref{eq:boundary} as follows: for right-moving modes ($v_{\rm rel} > 0$), we set $f(\mathbf k,z_-) = f^{\textrm{eq}}(\mathbf k,z_-)$ with $z_- < 0$ far in the broken phase, and for left-moving modes ($v_{\rm rel} < 0$), we set $f(\mathbf k,z_+) = f^{\textrm{eq}}(\mathbf k,z_+)$ with $z_+ > 0$ far in the unbroken phase.  These split boundary conditions are required on physical grounds: the collision term equilibrates the density matrix in the positive time direction, which for right(left)-going modes is the positive (negative) $z$ direction. So we have only  to impose equilibration as a boundary condition in the negative time direction, i.e. negative (positive) $z$ for right-(left-)moving modes. The fact that right(left)-going modes equilibrate again at late time, for $z \to z_+$ ($z \to z_-$), provides a non-trivial check on our numerics.

Within the baseline model parameters in Table~\ref{tab:baseline}, we have performed 
a number of stability checks against different choices of $k_{\rm max}$, $N_k$, $N_\vartheta$, $z_{\pm}$. 
We required that quantities of physical relevance, e.g.~$n_{L,R}(z)$, remain stable at the percent level. 
We find that  $|z_{\pm}| = 5000/T$ and  $k_{\rm max} = 8 T$ are acceptable values.~\footnote{Decreasing $y_{L,R}$ and/or $g_*$ 
increases $L_{\rm coll}$, thus requiring larger values of $|z_{\pm}|$.}   
Moreover, binning as coarse as $N_k = 4$, $N_\vartheta = 6$ produces stable charge density profiles. 
Our results below have $N_k=N_\vartheta=8$. 

Given our discretized solutions for $f(\mathbf k,z)$ and $\bar f(\mathbf k,z)$, we compute the charge currents and densities in Eq.~\eqref{eq:currentdef} by converting the continuous integrals into discretized sums in the usual way: $\int d^3 k/(2\pi)^3 \to k_{\rm max}/(2\pi^2 N_k N_\vartheta)\sum_{k,\cos\vartheta} k^2 $.

\subsection{Numerical results}
\label{sec:results}

\subsubsection{Distribution functions}

\begin{figure}[!t]
\begin{center}
\mbox{\hspace*{-0.5cm}\epsfig{file=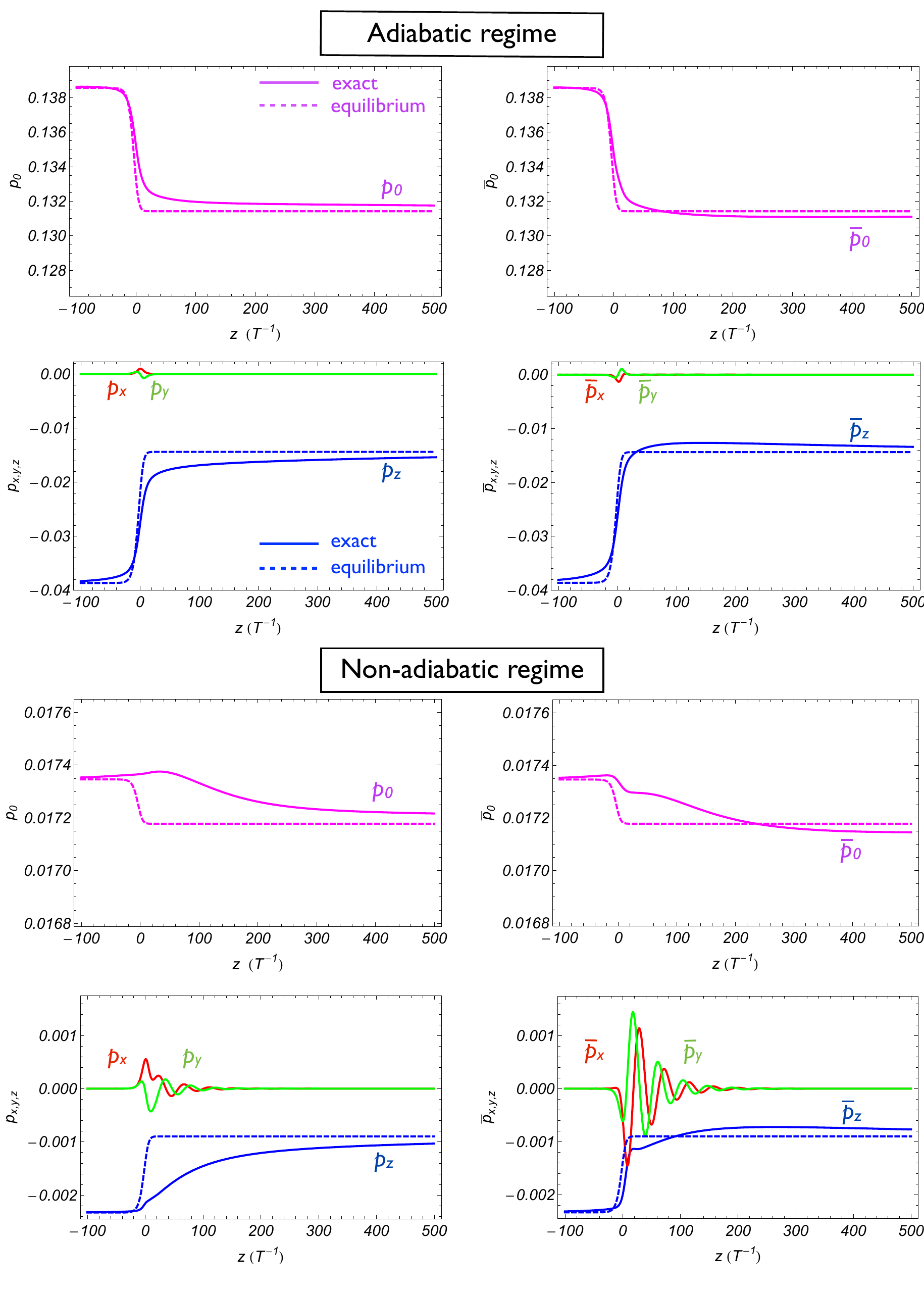,width=14.5cm}}
\end{center}
\vspace{-.75cm}
\caption{\it\small 
Numerical results for  particle and anti-particle density matrix 
for [top] a typical adiabatic bin: $k/T = 0.5, \cos\vartheta_k = 0.875$, and [bottom] a typical non-adiabatic bin, $k/T = 3.5, \cos\vartheta_k = 0.875$.
Left panels display particle density matrix in Bloch decomposition: $p_{0} (k,\cos\vartheta_k,z)$ and  $p_{x,y,z} (k,\cos\vartheta_k,z)$. 
Right  panels display anti-particle density matrix: 
$\bar{p}_{0} (k,\cos\vartheta_k,z)$   and  $\bar{p}_{x,y,z} (k,\cos\vartheta_k,z)$. 
Solid lines represent  full numerical solutions, while  
dotted lines represent local thermal equilibrium results for  diagonal components. 
See text for additional details. 
}
\label{fig:plop}
\end{figure}

Numerical results within the baseline model defined in Table~\ref{tab:baseline}  are presented in Fig.~\ref{fig:plop}. 
The figures show 
the $z$ dependence of $p_{0,x,y,z}$  (left panels) and  and $\bar{p}_{0,x,y,z}$ (right panels). 
The top four panels correspond to a typical  adiabatic bin 
with $k/T = 0.5, \cos\vartheta_k = 0.875$ and $L_{\rm osc}(z=0)/L_{w} \simeq 0.16$, while 
the bottom four panels correspond to a typical  non-adiabatic bin 
with $k/T =3.5, \cos\vartheta_k = 0.875$ and $L_{\rm osc}(z=0)/L_{w} \simeq 1.37 $.  ~\footnote{
In general  the  adiabaticity $L_{\rm osc}/L_{w}$  is controlled by $k, \cos\vartheta_k$, $m_{1,2}$, and $L_{w}$.  
Larger values of $k$ and $|\cos\vartheta_k|$  and smaller mass splittings increase $L_{\rm osc}$, 
thus leading to increasingly non-adiabatic evolution for fixed $L_{w}$. 
Once the underlying model parameters have been fixed, 
the adiabaticity is controlled only by  $k$ and $\cos\vartheta_k$.}
In all plots the solid lines represent the full numerical solutions, while the 
dotted lines represent local thermal equilibrium results. 

As anticipated, the qualitative behavior of Fig.~\ref{fig:plop}  can be readily understood 
through the magnetic analogy.  Let us neglect for a moment the effect of interactions with the thermal bath. 
In the  collisionless limit  different momentum bins are decoupled: 
$p_0$ and $\bar{p}_0$  do not evolve,
 while $\vect{p}$ and $\vect{\bar{p}}$ precess about effective magnetic fields 
as per Eqs.~(\ref{eq:precession1}).  Equilibrium boundary conditions for left- and right-moving modes 
imply that for $|z| \gg L_{w}$   both  $\vect{p}$ and $\vect{\bar{p}}$ point along the $\hat{\vect z}$ axis in flavor space.
This is a stable configuration  as long as $\vect{B}_{\Sigma} = 0$.  In proximity of the phase boundary, 
the non-vanishing $\vect{B}_\Sigma$ tends to push $\vect{p}$  and $\vect{\bar{p}}$ 
out of  their  stationary state, triggering the precession around the 
$z$-dependent  fields   $\vect{B}_0 \pm  \vect{B}_\Sigma$.
 
In the adiabatic regime ($L_{w}  \gg  L_{\rm osc}$), the polarization vectors   $\vect{p}$  
and  $\vect{\bar{p}}$   effectively track   the magnetic fields  $\vect{B}_0 \pm  \vect{B}_\Sigma$ 
(with a small precession amplitude that vanishes in the $L_{\rm osc}/ L_{w}  \to 0$ limit).   
As a consequence the solution tracks very closely the local thermal equilibrium.
On the other hand, in the non-adiabatic regime ($L_{w}  \leq L_{\rm osc}$),   
when the magnetic field changes on length scales comparable to or smaller  than 
the oscillation scale,  the polarization vector lags behind the magnetic field and begins  precessing with a large amplitude.  (In absence of collisions  the precession  persists  away from the phase boundary,  $|z| \gg L_{w}$.) 
The amplitude of oscillations increases with $L_{\rm osc}/L_{w}$,  as is evident from  Fig.~\ref{fig:plop}:  
in the non-adiabatic regime the system is  pushed out of equilibrium more efficiently by the passage of the bubble wall. 
Collisions and pair processes play an essential role in relaxing the 
density matrices back to equilibrium away from the phase boundary, 
as evident from the plots in Fig.~\ref{fig:plop}.

In the non-adiabatic regime,  CP-violating effects  show up more prominently in the evolution of the density matrices.  
In the CP-conserving limit ($\alpha'(z) = 0$), the effective magnetic fields 
$\vect{B}_0 \pm  \vect{B}_\Sigma$ are confined to a plane defined by $\hat{\mathbf z}$ and $(\sin\alpha \, \hat{\mathbf x}\!-\!\cos\alpha \, \hat{\mathbf y})$. The evolution obeys the CP invariance condition
$f(k,\cos\vartheta_k,z) = e^{i \eta} \bar{f}^T(k,\cos\vartheta_k,z) e^{-i \eta}$, where $\eta=\diag(\alpha,-\alpha)/2$, or
\be 
\label{eq:CPcond-polarization}
\bar{p}_x - i \bar{p}_y = e^{-i \alpha} \, (p_x + i p_y) \, , \quad
\bar{p}_z   = p_z  \, , \quad 
\bar{p}_0   = p_0   \; . 
\ee 
In presence of CP violation ($\alpha'(z) \neq 0$),   $\vect{B}_0 \pm  \vect{B}_\Sigma$ are not confined to this plane,
so that the dynamical evolution leads to an
angle between  $\vect{p}$  and $\vect{B}_0 + \vect{B}_\Sigma$ different from 
that between   $\vect{\bar{p}}$  and $\vect{B}_0 - \vect{B}_\Sigma$.
This leads to a violation of the conditions~(\ref{eq:CPcond-polarization}) and generation of 
flavor-diagonal CP asymmetries $p_z - \bar{p}_z \neq 0$  
(and eventually, through collisions, $p_0 - \bar{p}_0 \neq 0$).  
The  CP asymmetries vanish in two limits: (i) $L_{\rm osc}/L_{w} \ll 1$, because during the resulting adiabatic evolution the polarizations track closely the magnetic fields and so end up in CP-symmetric thermal equilibrium, and (ii) $L_{\rm osc}/L_{w} \gg 1$, because, as discussed in Ref.\cite{Cirigliano:2009yt},  
then the magnetic field varies so fast that precession becomes sensitive only to the initial and final values of $\vect{B}_0 \pm  \vect{B}_\Sigma$, which define a plane.
CP asymmetries are  maximal 
for $L_{\rm osc}/L_{w} \sim O(1)$.  

\begin{figure}[!t]
\begin{center}
\mbox{\hspace*{-1cm}\epsfig{file=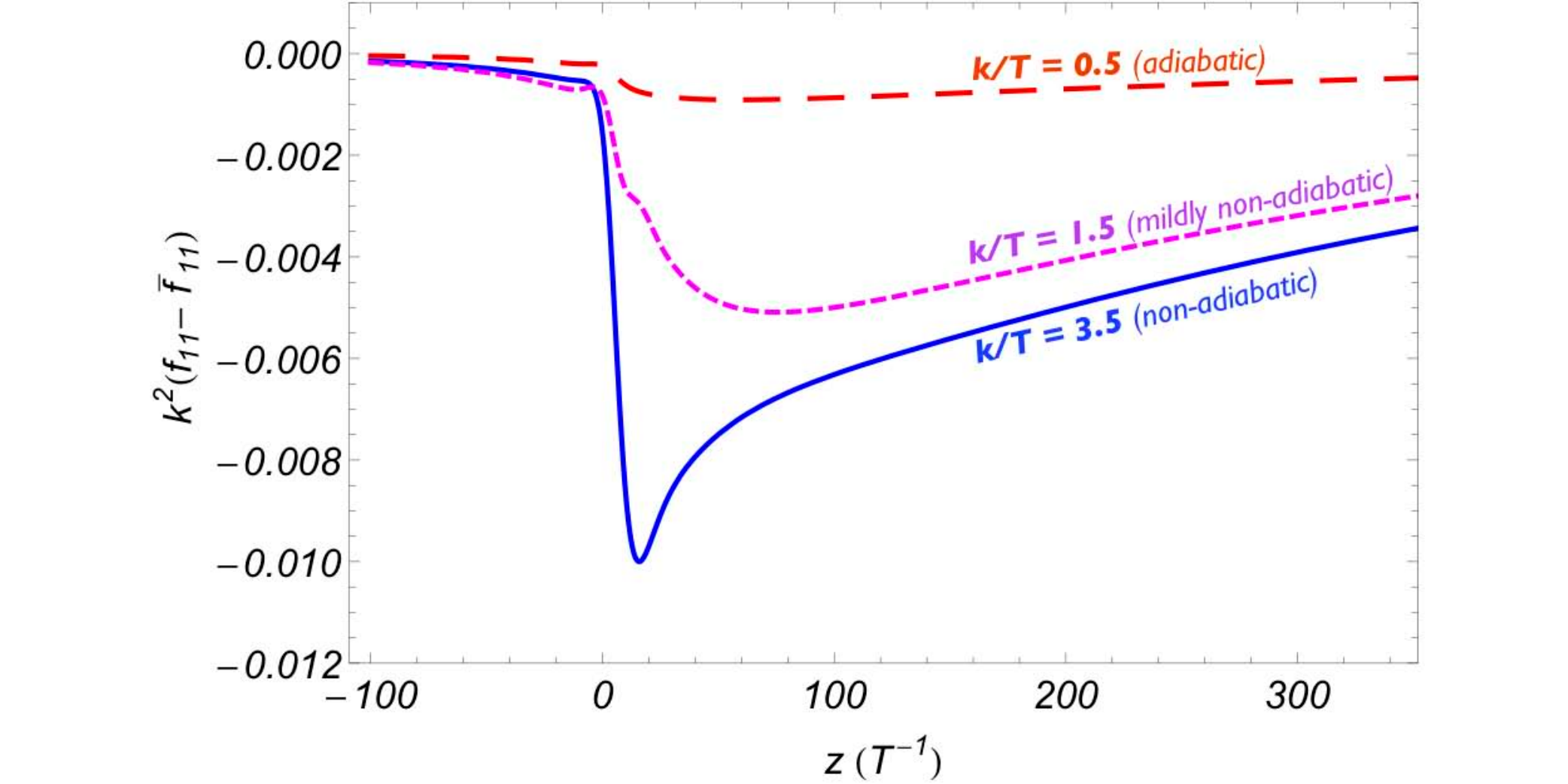,width=17cm}} 
\end{center}
\vspace{0cm}
\caption{\it\small 
Diagonal CP asymmetries $k^2 (f_{11} (z)  - \bar f_{11} (z))$ (in units of  $\,T^2$), 
for three representative bins.
The long-dashed line refers to a typical  adiabatic bin 
$(k/T = 0.5, \cos\vartheta_k = 0.875)$,  the short-dashed line to a mildly non-adiabatic bin
$(k/T = 1.5, \cos\vartheta_k = 0.875)$,  and the solid line to a typical non-adiabatic bin
$(k/T = 3.5, \cos\vartheta_k = 0.875)$.
}
\label{fig:cpz}
\end{figure}

In Fig.~\ref{fig:cpz}, we illustrate the ``anatomy'' of how CP asymmetries are sourced for different momentum bins.  Since $L_{\textrm{osc}} \sim 2 \pi k/|m_1^2 - m_2^2|$, we expect greater asymmetries to be generated for non-adiabatic bins corrsponding to larger values of $k$.  We plot the diagonal $\phi_1$ CP asymmetry $k^2 (f_{11}  - \bar{f}_{11} )$, 
for three representative $(k/T,\cos\vartheta_k)$ bins:
the long-dashed line refers to a typical adiabatic bin 
$(0.5, 0.875)$,  the short-dashed line to a mildly non-adiabatic bin
$(1.5, 0.875)$,  and the solid line to a typical non-adiabatic bin
$(3.5, 0.875)$.
We weight each asymmetry by the phase space factor $k^2$,  
so that Fig.~\ref{fig:cpz}  represents the contributions of each bin 
to the total charge density $n_1 =  \int d^3 k  (f_{11}  -  \bar{f}_{11})/(2\pi)^3$. 
The plots clearly illustrate:
\begin{enumerate}
\item[(i)] In the vicinity of the wall ($z \lesssim L_w = 20/T$), 
the largest asymmetry is generated for non-adiabatic momentum bins. 
\item[(ii)] Collisions establish kinetic equilibrium away from the wall $(L_w \lesssim z \lesssim 100/T)$, by redistributing 
charge among bins.  Far from the wall $(z \gtrsim 100/T)$, the density matrices are well-described by equilibrium 
distribution functions with a non-zero, spacetime-dependent chemical potential. 
\end{enumerate}
This picture holds 
as long as  $L_{\rm coll} >  L_{w}$, which is verified in the baseline model.  
A qualitative difference would arise in the case in which  $L_{\rm coll} \leq L_{w}$.
In that case we expect a suppression of the CP asymmetries~\cite{Cirigliano:2009yt}, because collisions are so frequent that they 
break the coherent evolution needed for a manifestation of CP-violating effects:
then flavor oscillations cannot play a significant role in generating a CP asymmetry.

\subsubsection{Charge densities and currents} 

\begin{figure}[!t]
\begin{center}
\mbox{\hspace*{-1cm}\epsfig{file=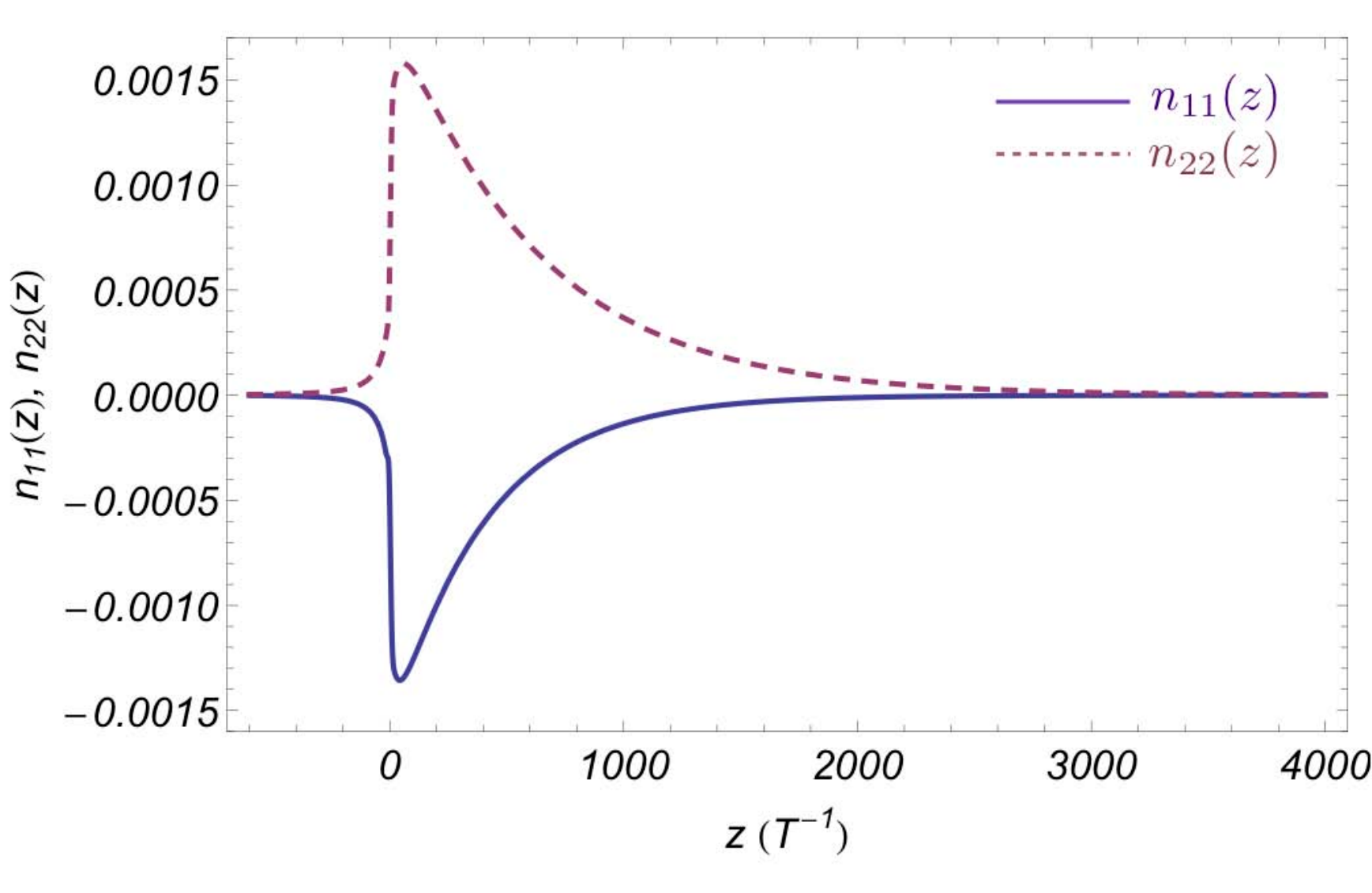,width=12cm}} 
\end{center}
\vspace{-.5cm}
\caption{\it\small 
Numerical results for the net charge densities  $n_{1} (z)$  (solid line) 
and $n_{2}(z)$ (dashed line), in units of $\, T^3$, 
within the baseline model specified in Table~\ref{tab:baseline}. 
In the unbroken phase ($z>L_{w}$)  $n_{1}=n_L$ and $n_{2}=n_{R}$  (mass and flavor basis coincide). 
This plot illustrates the existence of diffusion tails in the unbroken phase. 
The  relative size of the diffusion tails for  $n_{1}$ and $n_{2}$ is as expected, given that $y_L > y_R$ in the baseline model.  
}
\label{fig:densities}
\end{figure}

So far we have presented results for the mass-basis density matrices $f, \bar{f}$ (in the Bloch representation). 
In applications to EWBG,  one is interested in the behavior of flavor-diagonal  CP-violating charge densities 
$n_{L,R}(z)$ in the unbroken phase.
In Fig.~\ref{fig:densities}  we present numerical results  for the densities  $n_{1} (z)$ and $n_{2} (z)$.
(For $z\gg L_{w}$ one has $n_{L} = n_{1}$ and $n_{R} = n _{2}$.)
The plot in Fig.~\ref{fig:densities} clearly illustrates the existence of diffusion into the unbroken phase. 
Once generated by CP-violating oscillations  within the bubble wall, $\Phi_{L,R}$ charge diffuses into the 
unbroken phase, where mass and flavor eigenstates coincide and flavor oscillations no longer occur. 
The  smaller diffusion tail in  $n_{1} = n_L$ compared to $n_{2} = n_R$ is due to the  
fact that  $y_L > y_R$ in the baseline model, and so the mean free path for $\Phi_L$ is shorter.
On the other hand, in the broken  phase where flavor and mass  eigenstates
do not coincide,  flavor-sensitive collisions ($y_L \neq y_R$) 
lead to fast  flavor equilibration ($n_L - n_R  \to 0$). Since total charge conservation and 
causality imply $n_L + n_R \to 0$ far from the wall,  flavor equilibration has the effect of 
driving both diagonal densities  to zero~\footnote{This has essentially the same effect as the $\Gamma_m$ rates 
introduced in the diffusion equation treatment of this problem~\cite{Cohen:1994ss,Joyce:1994zn,Huet:1995sh}, 
although the physical mechanisms are not identical.}. 

Another interesting dynamical question involves the onset of the diffusion regime. 
Within our model we can evaluate the current densities 
and check whether there is a regime in which they satisfy the diffusion ansatz  
$\vect{j}_{ii}(z) =  -  D_{i}  \grad n_{i}(z)$,  with diffusion constants $D_{i}$.  
We have found that this ansatz is fairly well satisfied for large $z$ far from the wall ($z\gtrsim 300/T$) for some numerically fitted constant value for $D_{i}$. Of course, in principle one should calculate the diffusion coefficients from the collision terms. 
Nevertheless, this observation implies  that a simplified treatment in terms of diffusion equations for flavor-diagonal densities   
with appropriate  ``oscillation-induced"  sources  might lead to satisfactory results. 
We leave a more detailed investigation of this issue, including calculation of the diffusion constants in this model, to a forthcoming paper.

\subsubsection{Resonant enhancement of CP asymmetry}

Up to this point we have presented results for one  particular point in parameter space, 
defined by the baseline model (Table~\ref{tab:baseline}). 
In  phenomenological applications to baryogenesis, one would like to identify 
those regions of parameter space in which the CP asymmetries (and eventually the baryon asymmetry) are maximized. 
To this end, a useful global measure of CP violation 
and  a proxy for the baryon asymmetry itself within the toy model is 
provided by the integral $I_{L}^{CP}$ [see Eq.~\eqref{eq:ILR}], which represents the  
total $\Phi_L$ charge in the unbroken phase.

\begin{figure}[!t]
\begin{center}
\mbox{\hspace*{-1cm}\epsfig{file=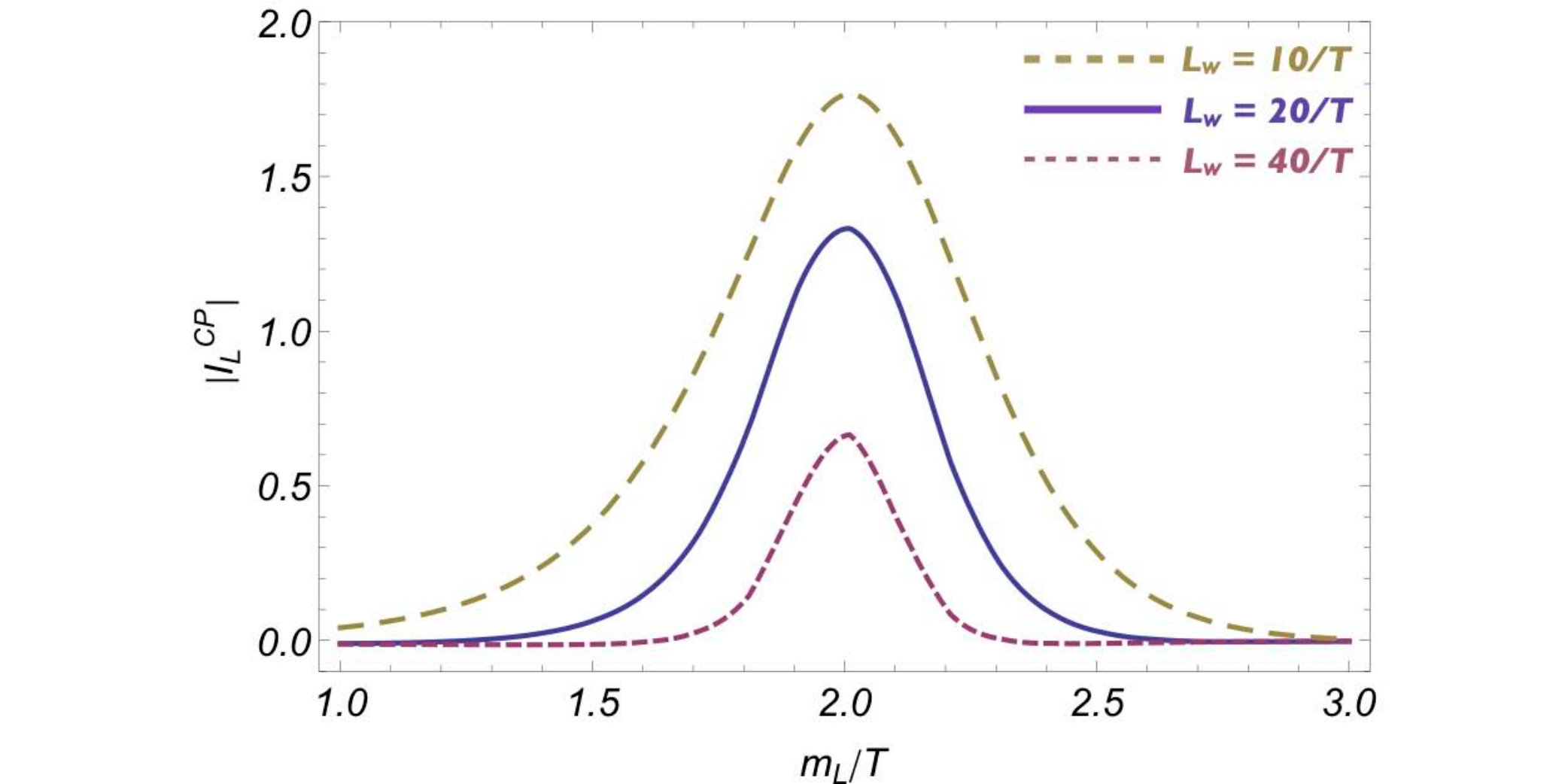,width=15cm}} 
\end{center}
\vspace{0cm}
\caption{\it\small 
Resonant enhancement of  $I_L^{CP}$, the total integrated $\Phi_L$ charge asymmetry in the unbroken phase [see Eq.~\eqref{eq:ILR}], as a function of 
the mass parameter $m_L$ (with fixed $m_R/T=2$) 
for different values of the wall thickness $L_{w}$:
$L_{w} = 10/T$ (long-dashed  curve), 
$L_{w} = 20/T$ (solid curve), 
$L_{w} = 40/T$ (short-dashed curve).
All other input parameters are as in  Table~\ref{tab:baseline}.}
\label{fig:resonance}
\end{figure}

From the discussion in the preceding section we have learned that the size of the CP asymmetry is 
controlled by the ratio $L_{\rm osc}/L_{w}$ of oscillation length to wall thickness.   
Here one should really think about  $L_{\rm osc}$ as a  thermally averaged 
oscillation length, or the oscillation length  evaluated at a  typical thermal momentum $k \simeq 3 T$. 
For fixed $v_0$, $L_{\rm osc}$ is controlled by the mass splitting $m_L - m_R$.  
In Fig.~\ref{fig:resonance}, we plot $I_L^{CP} $ versus  $m_L/T$,  for fixed 
$m_R/T=2$ and all other parameters as in Table~\ref{tab:baseline}. 
The dramatic resonant  feature at $m_L = m_R$ is interpreted in terms of non-adiabatic dynamics 
discussed in the previous section: for $m_L \sim m_R$ the average oscillation length $L_{\rm osc}$ is 
maximized,  implying that more momentum modes evolve non-adiabatically and therefore 
develop larger CP asymmetries.  
Fig.~\ref{fig:resonance} demonstrates in a consistent framework of  flavor mixing 
the resonant baryogenesis regime  previously discussed 
in the context of perturbative  mass-insertions~\cite{Carena:1997gx,Riotto:1998zb,Lee:2004we,Cirigliano:2006wh}
or perturbative insertions  of mass gradients~\cite{Carena:2000id,Carena:2002ss,Balazs:2004ae}, and places the origin of this resonance on a firmer theoretical footing.

Finally,  one can also study the dependence of the resonant enhancement 
of $I_L^{CP}$  on  other model parameters,  such as the wall velocity 
$v_w$, the coupling constant $y_L$, and the  wall thickness $L_{w}$. 
Decreasing (increasing) $v_w$ and $y_L$ increases  (decreases)  the size of the diffusion tail in $n_{L}(z)$, 
and hence leads to a larger (smaller)  $I_L^{CP}$.
The dependence on $L_{w}$ is more subtle than a simple overall scaling, because it affects both the 
peak and width of the resonance, as illustrated in Fig.~\ref{fig:resonance}. 
The resonance width is determined by the condition $L_{\rm osc}/L_{w} \gtrsim  {\mathcal O} (1)$. 
Numerically we find  considerable resonant enhancement for $|m_L - m_R| \lesssim  10/L_{w}$ 
(see Fig.~\ref{fig:resonance}). 
Moreover, as $L_{w}$ changes, the number of momentum modes that evolve non-adiabatically also changes, 
thus changing the overall peak of the resonance (decreasing $L_{w}$ leads to larger peak value for $I_L^{CP}$).

\section{Comparison with previous approaches}
\label{Sect:comparison}

In attempting to obtain a more tractable, analytic solution, all previous treatments have employed certain approximations to decouple the diagonal and off-diagonal components of the kinetic equations. In this section, we provide a detailed comparison of our results with the work of Ref.~\cite{konstandin}. That work provided the first derivation of the coupled two-flavor kinetic equations using the gradient expansion and treatment of flavor oscillations in EWBG.  Earlier works neglected quantum coherence implicitly by projecting onto diagonal densities within a diffusion-type ansatz.  

Our treatment and that of Ref.~\cite{konstandin} differ at the stage of power counting and solving these equations. The two primary differences are:
\begin{itemize}
\item \emph{Diagonal densities:} The power counting of Ref.~\cite{konstandin} leads one to neglect the diagonal components of the source $[\Sigma,f]$ so that $f_{11,22}$ do not depart from equilibrium at first order in $1/(L_w T)$. Effectively, this prevents any CP-asymmetry in the flavor-diagonal densities $n_{L,R}$  
(generated in the bubble wall) from diffusing into the unbroken phase. In contrast, in our treatment we find that in a consistent power counting scheme  deviations of $f_{11,22}$ from equilibrium \emph{are} sourced by nonzero $f_{12}$, and then diffuse deep into the unbroken phase, where the mass and flavor bases coincide. 

\item  \emph{Off-diagonal densities:} The off-diagonal density $f_{12}$ in Ref.~\cite{konstandin}  is sourced only by equilibrium diagonal densities, and 
its approach to equilibrium is described with a phenomenological ansatz for the collision term. While in some regimes of parameter space 
these simplifications capture the qualitative behavior of $f_{12}$ fairly well, quantitatively they lead to $\mathcal{O}(1)$ deviations  from the exact $f_{12}$. 
In our treatment, we account  for all contributions to $f_{12}$ from the source and collision terms at leading nontrivial order in our power counting.
\end{itemize}
Ultimately, the approximations of Ref.~\cite{konstandin}, applied to our toy model, lead one to neglect diffusion and result in a substantial 
underestimation of charge in the unbroken phase, compared to our exact numerical treatment, as  illustrated dramatically below in Fig.~\ref{fig:comp2}.  The reasons for this discrepancy are explained in detail below.

\subsection{Source and Collision Terms}

The  coupling of the different components of the $2\times 2$ distribution functions $f(\vect{k},z)$ arises from two sets of terms. In the mass basis, one is the CP-violating source proportional to $\Sigma(z)$. The other is the collision term $\mathscr{C}$. Our treatment accounts for the full coupled structure of both terms, while Ref.~\cite{konstandin} argued that the evolution of diagonal and off-diagonal densities could be decoupled in the source and collision terms.

Beginning with the source term, consider the components of its matrix structure:
\be
\Big[ \Sigma, f \Big] 
=  
\left(
\begin{array}{cc}
\Sigma_{12} f_{21} - \Sigma_{21} f_{12}   & 
-  \Sigma_{12}  \left( f_{11} - f_{22}  \right) +   \left(\Sigma_{11} - \Sigma_{22} \right) f_{12}  
\\
& \\
\Sigma_{21}  \left( f_{11} - f_{22}  \right) -  \left(\Sigma_{11} - \Sigma_{22} \right) f_{21}  
 & 
 \Sigma_{21} f_{12} - \Sigma_{12} f_{21}  
\end{array}
\right) ~. 
\label{eq:matrix-coupling}
\ee
In the diagonal entries, we find that the off-diagonal distribution functions $f_{12,21}$ source the diagonal distributions $f_{11,22}$ through $\Sigma_{12,21}$. In the off-diagonal entries, $f_{11,22}$ feed back to act as sources for $f_{12,21}$. In our work, we have not made any a priori assumptions about the scalings of the $f$'s with gradients of the external field ($\ewall$) and thus accounted for the full coupled evolution. In Ref.~\cite{konstandin}, however, all deviations of $f_{ij}$ away from their equilibrium values were power counted as $\mathcal{O}(\ewall)$. That is,
\be
\label{eq:linearized}
\text{Ref.~\cite{konstandin}:}\quad f = 
\begin{pmatrix}
n_B(\omega_1) & 0 \\
0 & n_B(\omega_2) 
\end{pmatrix}
+ \begin{pmatrix}
\delta f_{11} & f_{12} \\
f_{21} & \delta f_{22}\,
\end{pmatrix},
\ee
where $\delta f_{11,22}$ and $f_{12,21} \sim \mathcal{O}(\ewall)$.  According to this counting, the source term is
\be
\Big[ \Sigma, f \Big]_{\text{Ref.~\cite{konstandin}}}
=  
\left(
\begin{array}{cc}
0  & 
-  \Sigma_{12}  \left( n_B(\omega_1) - n_B(\omega_2) \right) \\
& \\
\Sigma_{21}  \left( n_B(\omega_1) - n_B(\omega_2) \right)
 & 
 0
\end{array}
\right) \; + \mathcal{O}(\ewall^2) ~. 
\label{eq:decoupled-source}
\ee
Ref.~\cite{konstandin} argued that, working at $\mathcal{O}(\ewall)$, the $\mathcal{O}(\ewall^2)$ terms could be neglected.

Here is the crucial point: in the power counting of Ref.~\cite{konstandin}, there is {\it no} CP-violating source for the diagonal densities.  Furthermore, although there is a source for the off-diagonal densities, the different components of $f_{ij}$ are decoupled; CP violation in $f_{12,21}$ does not feed into $f_{11,22}$.  By rotating to the flavor basis, one has $n_{L,R} = \mp \sin 2\theta \int d^3 k \,  \textrm{Re}[(f_{12}- \bar{f}_{12})e^{i \alpha}]/(2\pi)^3$.  Charges $n_{L,R}$ vanish in the unbroken phase, since $\theta = 0$.  No diffusion exists.

Now we consider the collision term for $f_{12}$. The coupled structure of $\mathscr{C}$ was simplified in Ref.~\cite{konstandin} by making a simple phenomenological ansatz
\be
\label{collisionansatz}
\text{Ref.~\cite{konstandin}:}\quad\mathcal{C}_{12}(\vect{k},z) = -\Gamma_{12} f_{12}(\vect{k},z)
\ee
for some constant relaxation rate $\Gamma_{12}$, which was estimated to be $\sim \alpha T$ where $\alpha$ is the coupling strength of the dominant interaction of the species. As we  discuss in Appendix \ref{app:decoupling}, there is indeed a part of  the full collision term that takes the form of this  ansatz, although with a $\vect{k}$-dependent rate $\Gamma_{12}(\vect{k},z)$. We will consider in the Appendix the full set of contributions to the collision term, and the conditions under which the ansatz \eq{collisionansatz} may be justified.

\subsection{Power Counting of Off-Diagonal Solution}
\label{ssec:offdiagonal}

We now consider solutions for $f, \bar{f}$ in our toy model, following the procedure of Ref.~\cite{konstandin}, by making the assumptions described above for the source and collision terms.  One obtains the decoupled equations for the off-diagonal densities,
\begin{subequations}
\label{eq:approx} 
\begin{eqnarray}
\Big[  v_{\rm rel}    \partial_z  + i (\omega_1 - \omega_2)   + v_{\rm rel} ( \Sigma_{11} - \Sigma_{22})   
+ \Gamma_{12}
 \Big]  \ f_{12}  (\vect{k}, z) & = & 
 v_{\rm rel}  S_{12}  
\\ 
\Big[  v_{\rm rel}    \partial_z  - i (\omega_1 - \omega_2)   + v_{\rm rel} ( \Sigma_{11} - \Sigma_{22})     
+ \bar{\Gamma}_{12}
 \Big]  \ \bar{f}_{12}  (\vect{k}, z) & = & 
 v_{\rm rel}  {S}_{12}  ~. 
\end{eqnarray}
\end{subequations}
where 
\bea
\label{S12def}
S_{12} (\vect{k}, z) &=& 
\Sigma_{12} (z) \Big( n_B (\omega_{1\mathbf k}) - n_B(\omega_{2 \mathbf k}) \Big)  ~. 
\eea
The effective sources for particles and antiparticles coincide, as they only involve 
the diagonal equilibrium densities $f_{ii}^{\textrm{eq}} = \bar{f}_{ii}^{\textrm{eq}} = n_B (\omega_{i\mathbf k})$. For the relaxation rates $\Gamma_{12},\bar\Gamma_{12}$ in \eq{eq:approx}, we will adopt a  constant value  $\Gamma_{12} = \bar{\Gamma}_{12} = 0.024 \,  T$,  obtained by 
evaluating \eq{relaxationrate} at $k = 3T$ and $\cos\vartheta_k = 1$.\footnote{In general 
$\Gamma_{12} (\vect{k},z)$  in \eq{relaxationrate} varies considerably with $k$ (it decreases by a factor of four over the range $0 <  k/T < 9$) 
 but has a weak dependence on  $\vartheta_k$ and $z$.}

The simplified \eq{eq:approx} can be solved explicitly, giving~\cite{konstandin} 
\be
\label{decoupledf12solution}
f_{12}(\vect{k}; z) = \int_{-\infty}^z \! \! dz'  
\,  S_{12} (\vect{k}, z')  \,
 \exp\left\{- \!\! \int_{z'}^{z} \! \! dz'' \left[ i\frac{(\omega_{1\mathbf k}-\omega_{2\mathbf k})}{ v_{\text{rel}}}  +  (\Sigma_{11}-\Sigma_{22})+ \frac{\Gamma_{12}}{v_{\text{rel}}} 
 \right]\!(z'') 
\right\} .
\ee
One can notice from this solution, however, that power counting $f_{12}$ as $\mathcal{O}(\ewall)$ breaks down in the non-adiabatic regime. Now, it is true that $\Sigma_{12}$ which explicitly appears in the integrand of \eq{decoupledf12solution} is $\mathcal{O}(\ewall)$. However, the integration measure  also must be power counted. It is determined by the \emph{shortest} length scale among the wall length $L_{w}$, the oscillation length $L_{\rm osc}$ or the collision length $L_{\textrm{coll}} \sim 1/\Gamma_{12}$, since that will determine the effective range of integration. Supposing the latter to be very long, we can just compare $L_{w}$ and $L_{\rm osc}$. In the non-adiabatic regime ($L_{w}\lesssim L_{\rm osc}$) $L_{w}$ is the shortest scale; the effective range of integration in \eq{decoupledf12solution} is $L_{w}$ since $\Sigma_{12}(z')$ is nonzero only in the bubble wall region. However, we note that $L_{w}\sim 1/\ewall$ and $f_{12}$ in \eq{decoupledf12solution} should be  power counted as
\be
\label{eq:f12powercounting}
f_{12} \sim (\text{range of integration})\times \Sigma_{12} \sim \frac{1}{\ewall}\times \ewall = \ewall^0\,,
\ee
invalidating the estimate in Ref.~\cite{konstandin} that $f_{12}\sim \ewall$.\footnote{In our power counting scheme, $S_{12}$ in \eq{S12def} is also order $\eosc$, so one may ask if the terms with $f_{12,21}$ in the source \eq{eq:matrix-coupling} may be counted as order $\epsilon^2$ and be dropped, mimicking Ref.~\cite{konstandin}. However, as we show in Appendix~\ref{app:power}, if one power counts this way then \emph{every} term in the kinetic equation is order $\epsilon^2$, and every term in \eq{eq:matrix-coupling} must be kept, thus keeping the equations coupled.}

\begin{figure}[!t]
\begin{center}
\mbox{\hspace*{-0.5cm}\epsfig{file=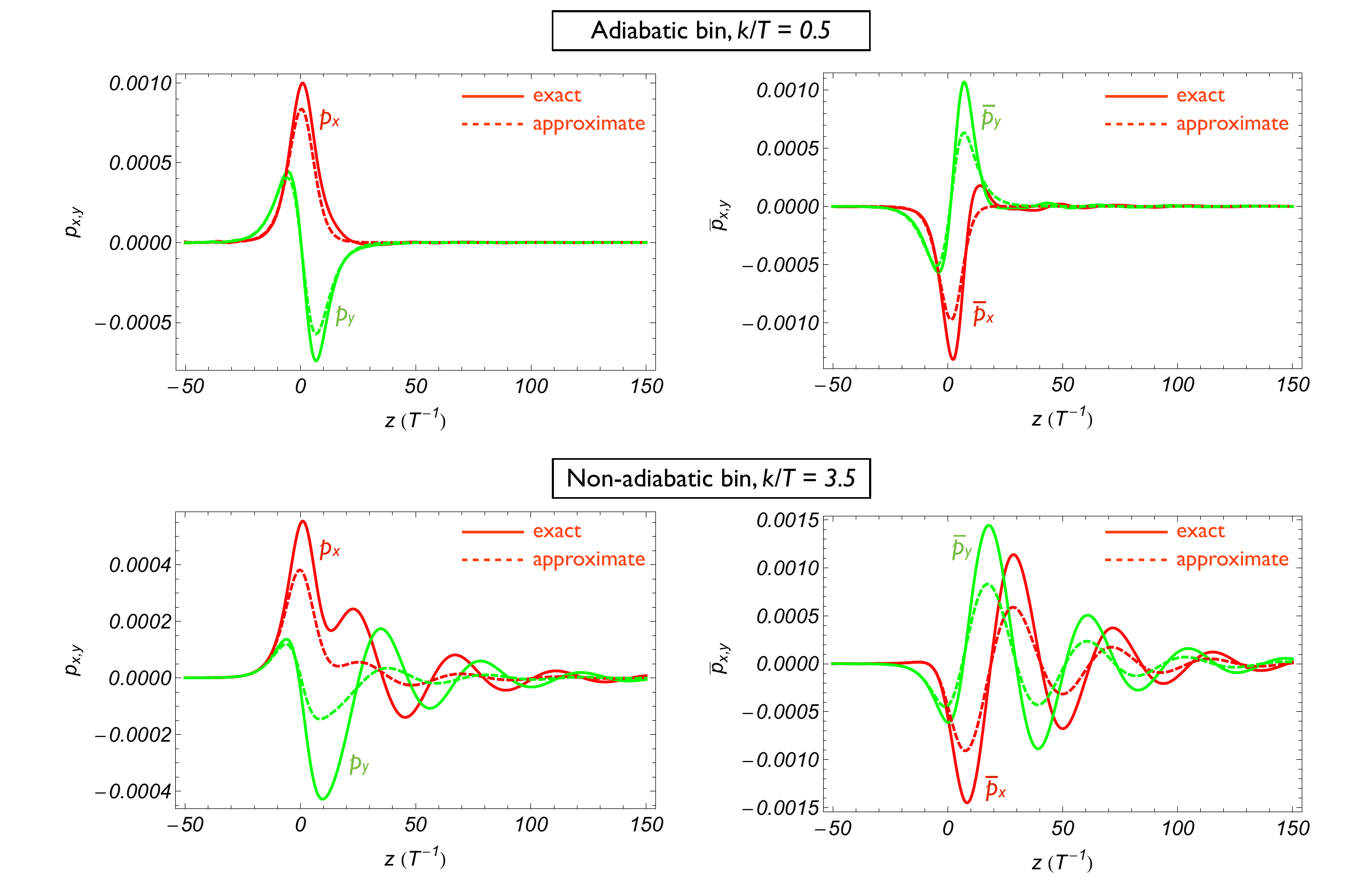,width=17.5cm}} 
\end{center}
\vspace{-.25cm}
\caption{\it\small 
Numerical results for the off-diagonal components  
$p_{x,y} (k,\cos\vartheta_k,z)$ 
of the  particle density matrix 
(left panels) and anti-particle density matrix
$\bar{p}_{x,y} (k,\cos\vartheta_k,z)$  (right panels).
Solid lines represent  solutions of the full equations. Dashed lines represent  
solutions  of the approximate equations (\ref{eq:approx}). 
The upper panels correspond to a  typical adiabatic bin with  $k/T = 0.5, \cos\vartheta_k = 0.875$, 
while the lower panels correspond to a typical non-adiabatic bin with  $k/T = 3.5, \cos\vartheta_k = 0.875$. 
}
\label{fig:comp1}
\end{figure}

In the adiabatic regime, the power counting of Ref.~\cite{konstandin} is formally consistent, as explained in Appendix~\ref{app:power}.  As a result, \eq{decoupledf12solution}, sourced only by the equilibrium diagonal distributions in \eq{S12def}, is actually a fairly good approximation for the true solution for $f_{12}$, as illustrated below in the top two panels of Fig.~\ref{fig:comp1}.

The above observations can be easily understood from Fig.~\ref{fig:precession}, which illustrated the magnetic analogy for precession of flavor polarization vectors. In equilibrium, $\abs{\vect{p}}$ starts out proportional to $n_B(\omega_1) - n_B(\omega_2) \sim \eosc$. Also the initial magnetic field $\vect{B}_0\sim\omega_1-\omega_2\sim \eosc$. When the wall $\vect{B}_\Sigma\sim \ewall$ turns on, the precession angle is thus of order $\theta_B\sim \ewall/\eosc$. In the adiabatic regime, this ratio is small, and so $\theta_B\ll 1$. Then, as $\vect{p}$ precesses, the off-diagonal deviations are given by $p_{x,y} \sim \abs{\vect{p}}\theta_B \sim (\eosc)(\ewall/\eosc)\sim\ewall$. Meanwhile, the deviation in the $z$-component giving the diagonal densities is $\delta p_z\sim \abs{\vect{p}}(1-\cos\theta_B) \sim \eosc\theta_B^2\sim \ewall^2/\eosc$. Thus we can neglect the feedback of $\delta p_z$ to $p_{x,y}$, and using the decoupling approximation to solve for $f_{12} = p_x + i p_y$ is justified.

However, in the non-adiabatic regime, $\ewall\gtrsim\eosc$, the precession angle is order 1. Then the deviations in $p_{x,y,z}$ are all the same order, $\delta p_{x,y,z}\sim \abs{\vect{p}}\sim \eosc$, and none of them can be neglected or approximated as being in equilibrium. The magnetic analogy makes clear why the decoupling approximation breaks down in the non-adiabatic regime---if the entire vector $\vect{p}$ precesses with a large angle, all components change simultaneously with equal magnitudes, and there is no notion of decoupled evolution of the individual components.

\subsection{Power Counting of Diagonal Solution}
\label{ssec:diagonal}

For the diagonal densities, we can power count deviations of $f_{ii},\bar f_{ii}$ from equilibrium as in \eq{eq:f12powercounting}. We find in the non-adiabatic regime, $\delta f_{ii}\sim \ewall^0$, so they certainly cannot be neglected. In the adiabatic regime,  $\delta f_{ii}\sim \ewall^2$. Then it appears that the power counting of \cite{konstandin} is justified in this  regime. However, this is not the case. 

First, deviations of diagonal distributions from equilibrium survive much farther into the unbroken phase ($z>0$) than off-diagonals, as we found in \fig{fig:plop}. In \fig{fig:cpz} we found that the part surviving to very large $z$ is well approximated by a chemical potential that survives even after kinetic equilibrium ($f_{ii} = n_B(\omega_i-\mu_i), f_{12} = 0$) is reached much earlier. This is because collisions drive $f_{ij}$ to kinetic equilibrium on a length scale $L_{\text{coll}}$, but  the chemical potential is damped away on a longer diffusion length scale $L_{\text{diff}} \sim L_{\text{coll}}/v_w$. Thus, for very large $z$, the diagonal $\delta f_{ii}$ should not be neglected even in the adiabatic regime, although  they are formally suppressed relative to $f_{12}$ in the region of the wall.

Second, as we argued in \sec{Sect:solution}, one is ultimately interested in the total integrated charge $I_{L,R}^{CP}$ diffusing into the unbroken phase, and the amount of this charge is governed by the diffusion length $L_{\text{diff}}$. In our model $L_{\text{diff}}$ is actually the largest length scale in the problem.  Thus, although in the adiabatic regime the deviations of $ f_{ii},\bar f_{ii}$ from equilibrium are formally $\mathcal{O}(\ewall^2)$, their contributions to $I_{L,R}^{CP}$ are order $\ewall^2/(\ecoll v_w)$ due to the measure of integration, and thus parametrically larger than na\"{i}vely expected. In other words, even in the adiabatic regime, one should not neglect deviations from equilibrium in the diagonal densities.

\subsection{Numerical Comparisons and Diffusion Tail}

In Fig.~\ref{fig:comp1}, we compare our exact solution for $f_{12}(\vect{k},z)$ and $\bar f_{12}(\vect{k},z)$ (solid lines) 
with the approximate solution \eq{decoupledf12solution} (dashed lines)  in a typical adiabatic bin and non-adiabatic bin, 
given in terms of the components $p_{x,y},\bar p_{x,y}$ in the Bloch decomposition \eq{Bloch}.
Overall, the qualitative behavior is similar.  Quantitatively,
the solutions disagree in the non-adiabatic bins, while they are somewhat closer in the adiabatic bin, consistently  with the power counting estimates given above. The remaining discrepancies may be due to effects missed by the ansatz \eq{collisionansatz} for the collision term, on which we  comment in Appendix~\ref{app:decoupling}, and due to neglecting the feedback of $\delta f_{11,22}$ to $f_{12}$ through the wall-induced source in \eq{eq:matrix-coupling}, on which we comment in Appendix~\ref{app:power}.

\begin{figure}[!t]
\begin{center}
\mbox{\hspace*{-0.25cm}\epsfig{file=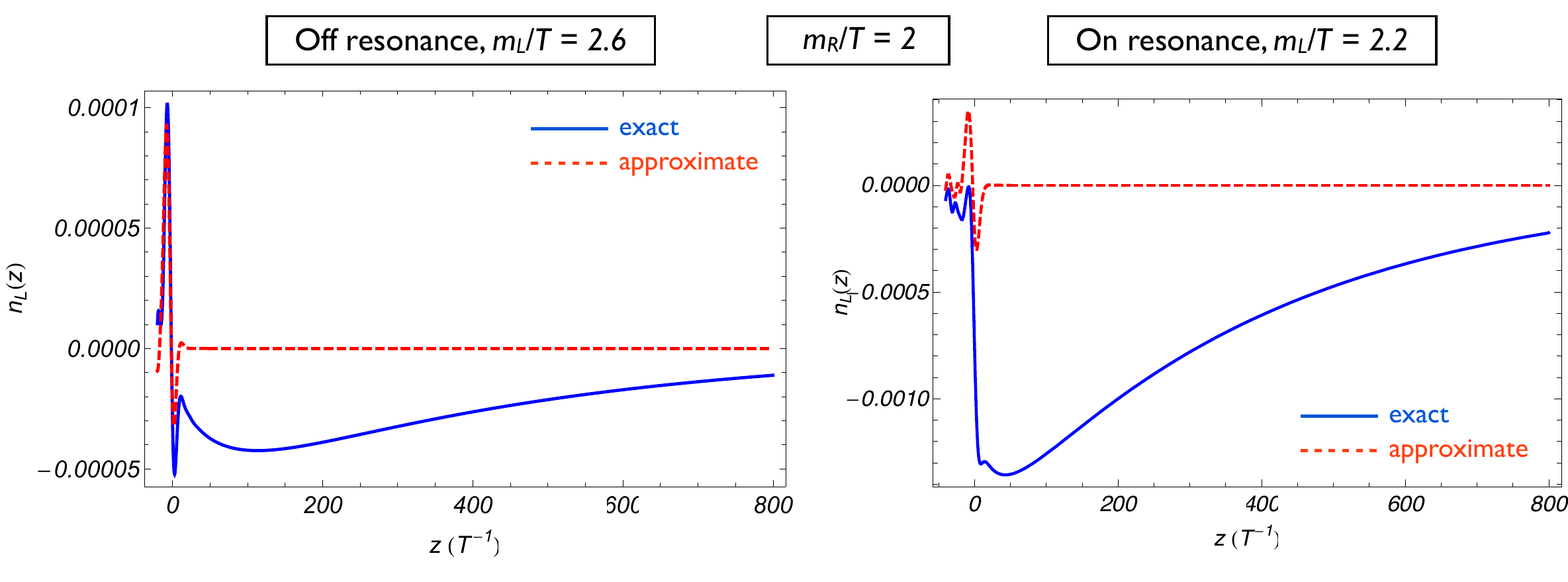 ,height=6cm}} 
\end{center}
\vspace{-1cm}
\caption{\it\small 
Charge density profiles $n_L (z)$ from the solution of the full equations (solid line) 
and the approximate decoupled  equations (dashed line) that mimic the procedure of 
Ref.~\cite{konstandin}. The left panel correspond to the off-resonance regime 
$m_L/T = 2.6, m_R/T = 2$, while the right panel corresponds to the 
resonant regime  $m_L/T = 2.2, m_R/T = 2$.
}
\label{fig:comp2}
\end{figure}

The most significant discrepancy, however, between the treatment of Ref.~\cite{konstandin} and ours arises in  
the diagonal distributions, and is illustrated in Fig.~\ref{fig:comp2}. Our exact solutions exhibit diffusion of the diagonal flavor densities $n_{L,R}$ deep into the unbroken phase (where they are equal to $n_{1,2}$), while this phenomenon is absent in the treatment of Ref.~\cite{konstandin}. The reason is that the decoupling imposed in the source term in \eq{eq:decoupled-source} does not allow the off-diagonal densities $f_{12,21}$ to source the diagonals 
$f_{11,22}$. Thus, in  Ref.~\cite{konstandin}, the flavor-diagonal densities are obtained simply by rotating $f_{12,21}$ back to the flavor basis. 
But this generates nonzero $n_{L,R}$ only \emph{inside} the wall, not outside, where the mass and flavor bases coincide. 
On the other hand  we showed above  that in a consistent power counting the fully-coupled source \eq{eq:matrix-coupling} must be kept. Once the diagonals $f_{11,22}$  are sourced, through  collisions they diffuse outside the wall into the unbroken phase. 
This is the  main mechanism to generate $n_L$ in the unbroken phase.

\section{Conclusions}

Electroweak baryogenesis is an attractive and testable explanation for the origin of the baryon asymmetry of the Universe.  However, quantitative baryogenesis predictions for collider and low-energy precision measurements are obscured by orders-of-magnitude discrepancies between different treatments of the charge transport dynamics during the electroweak phase transition, due to different approximations.  Since the final baryon asymmetry depends on how much CP-asymmetric charge is generated and transported into the phase of unbroken symmetry, a quantitative computation of these dynamics is essential.

This work has provided an important step (beginning with Ref.~\cite{Cirigliano:2009yt}) toward investigating these discrepancies and providing a quantitatively robust formalism for baryogenesis computations.  
Within a generalized gradient expansion, we derived Boltzmann equations for the particle and antiparticle density 
matrices for a two-flavor scalar system with an inhomogeneous, CP-violating mass matrix.  (One may regard the scalar species as a toy model for squarks in supersymmetric extensions of the Standard Model, with an inhomogeneous mass matrix arising from the spacetime-varying Higgs field during the phase transition.)  Our Boltzmann equations are ``flavored'' in the sense that they keep track not only of the occupation numbers of individual 
states, but also of their coherence. 
These equations account for flavor oscillations in a non-homogeneous background 
and interactions with a thermal bath.  
In contrast to previous treatments, we have kept the full matrix structure of the collision term 
and have not resorted to the usual diffusion approximation.  However, it is clear that diffusion does emerge
from our full numerical solutions. 
From our analysis a very simple physical picture emerges: 
at leading order in gradients of the mass matrix,  
CP asymmetries arise from coherent flavor oscillations induced by spacetime-dependent mixing. 

By virtue of our simplified model, we solved the Boltzmann equations numerically without ansatz for the form of the density matrices.
We illustrated several important physics points: 
\begin{itemize}
\item The largest departures from equilibrium and the largest contributions to the CP-violating asymmetries 
arise for states evolving non-adiabatically across the phase boundary, with momenta ${\mathbf k}$ satisfying
$L_{\rm osc} = 2 \pi v_{\rm rel}(\mathbf k) / (\omega_{1\mathbf k} - \omega_{2\mathbf k})  \gtrsim   L_{w}$, where  $L_{w}$ is the bubble wall thickness (i.e. the length over which the off-diagonal elements of the mass matrix vary). 
\item  The CP-violating flavor-diagonal charge densities, generated by flavor oscillations within the wall, diffuse into the unbroken phase and are not localized near the wall.
\item  The enhancement of charge generation in the non-adiabatic regime 
manifests itself as a resonance when $m_L \sim m_R$.  This ``resonant regime'' is governed by the condition $|m_L - m_R| \lesssim 10 L_w^{-1}$ (i.e., the width of the resonance is controlled by $L_{w}$). In MSSM-like models, $L_w\sim 20/T$, so for mass differences as big as $\abs{m_L-m_R}\sim T/2$ or about $50\GeV$, it becomes important to account consistently for modes that evolve non-adiabatically, as we have done.
\end{itemize}
These findings demonstrate the crucial importance of keeping track of full coupled evolution of all components of the density matrix $f$ to capture the dominant contributions to flavor-diagonal charge densities that diffuse into the unbroken phase.

We also compared our results, within the context of our toy model, to the  formalism of Ref.~\cite{konstandin}.
All previous baryogenesis computations have relied on an implicit or explicit decoupling of the 
dynamics of diagonal and off-diagonal densities, whereas our  results here do not and achieve exact numerical solutions for the full density matrices. 
We have shown that the power counting of Ref.~\cite{konstandin} (leading to  decoupled equations for diagonal and off-diagonal densities)
is inadequate in the non-adiabatic regime, in which the CP-violating effects are maximal. 
We have solved our simple model of mixing scalars 
according to the procedure outlined in Ref.~\cite{konstandin},  
finding dramatic differences in the charge density profiles:  
the full solution displays significant diffusion into the unbroken phase that is absent in the approximate treatment.  
This difference can be directly traced to the approximation of decoupling the kinetic 
equations for diagonal and off-diagonal sources. 
This difference may have a potentially large impact on 
electroweak baryogenesis calculations. 
Our analysis indicates that the approach of 
Ref.~\cite{konstandin}  largely underestimates the CP-violating densities in the unbroken phase, 
which in turn might induce a large underestimation  (by one order of magnitude or more)  of the produced baryon asymmetry.
Within the simple model of mixing scalars, however, we cannot address this in a quantitative manner, 
as we still need to introduce fermions.  An additional CP-violating source, the semi-classical force~\cite{Cline:1997vk, Cline:2000nw, Kainulainen:2001cn,Kainulainen:2002th,Zhou:2008yd}, can arise for fermions, but not scalars.  The relative magnitude between this source and the resonant, mixing-induced source studied here remains an important open question.

The resolution of current discrepancies and a more robust phenomenological analysis require 
the following additional steps, which are currently under investigation:
\begin{itemize}
\item  Transport equations for fermions with an inhomogeneous mass matrix along the same lines as our analysis of mixing scalars, including the resonant mixing-induced source, the semi-classical force, and elastic and inelastic scattering processes.
\item Identification of diffusion equations with appropriate 
oscillation-induced sources that correctly capture the physics of the full kinetic analysis. 
This should be a good description of the system in the unbroken phase,  where the mixing 
angle vanishes and flavor oscillations no longer occur.  
\end{itemize} 
These  developments, building upon the methods we have introduced here and in Ref.~\cite{Cirigliano:2009yt}, will make possible rigorous and tractable predictions for charge transport in realistic scenarios of EWB.

\begin{acknowledgments}

We would like to thank Michael Ramsey-Musolf for collaboration in earlier stages of this work and many helpful comments that improved this paper. We thank Matti Herranen, Kimmo Kainulainen, Thomas Konstandin, and Tomislav Prokopec for many insightful discussions.
The work of VC is supported by the Nuclear Physics Office of the U.S.
Department of Energy under Contract No.~DE-AC52-06NA25396 and by the LDRD program at Los Alamos National Laboratory. The work of CL is supported by the U.S. Department of Energy under Contract No. DE-FG02-94ER40818. 
ST is supported by the NSERC of Canada and would like to thank B. Garbrecht for teaching him the method of relaxation.

\end{acknowledgments}


\appendix

\section{Collision terms}
\label{app:coll}

We evaluate the interaction terms in the Boltzmann equations coming from $\Lint$, given by Eq.~\eqref{eq:lint}.  We work in the mass basis below, but omit the subscripts $m$ for brevity.  We assume that the $A$ bosons (mass $m_A$) are in thermal equilibrium, with temperature $T \gg m_A$.    

\begin{figure}[t!]
\begin{center}
\mbox{\hspace*{-1cm}\includegraphics{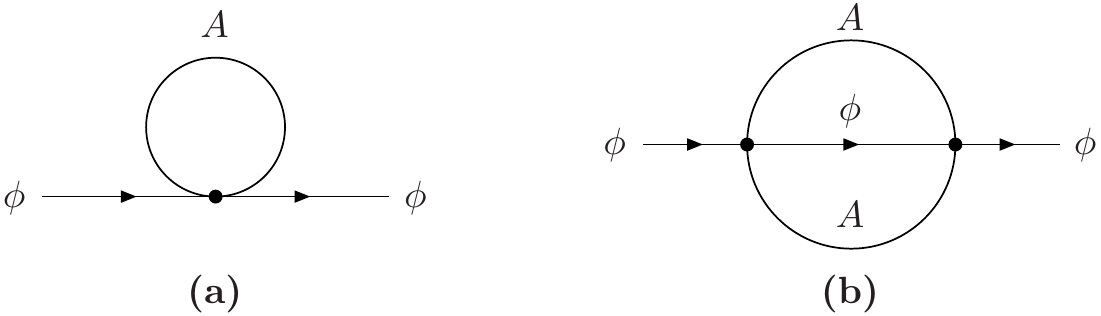}}
\end{center}
\vspace{-.5cm}
\caption{\it\small 
Leading-order self-energy graphs that induce the collision terms in the Boltzmann equations, corresponding to (a) thermal mass corection from coherent foward scattering, and (b) non-forward scattering ($\phi A \leftrightarrow \phi A$) and annihilation ($\phi \phi^\dagger \leftrightarrow A A$).
}
\label{fig:feyn}
\end{figure}

The self-energy functions $\Pi$ can be computed following a perturbative expansion in $y_{L,R}$, detailed in Ref.~\cite{Cirigliano:2009yt}.  At linear order in $y_{L,R}$, only the $[\Pi^h, G^\gtrless]$ term receives a contribution, shown by Fig.~(\ref{fig:feyn}a):
\be
i \Pi^h(k,x) = Y(x) \, \left( \frac{T^2}{24} + \int \! \frac{d^3 p}{(2\pi)^3} \frac{1}{ 4\epsilon_{\mathbf p}} \right) \; .
\ee
The first term gives a thermal mass shift: $m_{L,R}^2 \to m_{L,R}^2 + y_{L,R} T^2/24$. The second term ($\epsilon_{\mathbf p} \equiv \sqrt{p^2 + m_A^2}$)  is the usual zero-temperature divergence that can be absorbed by renormalization.

At second order in $y_{L,R}$, scattering and annihilation processes arise from the imaginary part of Fig.~(\ref{fig:feyn}b) and are given by Eq.~\eqref{colldef}.  The scattering term $[\phi(k) A(p) \leftrightarrow \phi(k^\prime) A(p^\prime)]$ is given by 
\begin{align}
&\mathscr{C}[f, \bar f]_{\textrm{scat}} = \; - \; \frac{1}{4 k^0}  \int\!\! \frac{d^3 k^\prime}{(2\pi)^3 2 k^{\prime 0}} \int\!\! \frac{d^3 p}{(2\pi)^3 2\epsilon_{\mathbf p}} \int\!\! \frac{d^3 p^\prime}{(2\pi)^3 2\epsilon_{\mathbf p^\prime}} \; (2\pi)^4 \, \delta^4(k+p-k^\prime - p^\prime) \label{Cscat} \\
&\; \times \left[  \bigl\{ f(\mathbf k) , \, Y(1 + f(\mathbf k^\prime))Y  \bigr\}f_A(\mathbf p)(1+ f_A(\mathbf p^\prime)) -  \bigl\{ (1+f(\mathbf k)), \, Yf(\mathbf k^\prime) Y  \bigr\}(1+f_A(\mathbf p))f_A(\mathbf p^\prime)  \right] \notag 
\end{align}
where $f_A(\mathbf p) \!=\! n_B(\epsilon_{\mathbf p}) \equiv 1/(\exp(\epsilon_{\mathbf p}/T) -1 )$ is the distribution function the $A$ bosons.  The annihilation term \mbox{$[\phi(\mathbf k) \phi^\dagger(\mathbf k^\prime) \leftrightarrow A(\mathbf p) A(\mathbf p^\prime)]$} is given by
\begin{align}
&\mathscr{C}[f, \bar f]_{\textrm{ann}} = \; - \; \frac{1}{8 k^0 } \int\!\! \frac{d^3 k^\prime}{(2\pi)^3 2 k^{\prime 0}} \int\!\! \frac{d^3 p}{(2\pi)^3 2\epsilon_{\mathbf p}} \int\!\! \frac{d^3 p^\prime}{(2\pi)^3 2\epsilon_{\mathbf p^\prime}} \; (2\pi)^4 \, \delta^4(k+k^\prime -p - p^\prime) \label{Cann}\\
& \times \left[  \bigl\{ f(\mathbf k) , \, Y \bar f(\mathbf k^\prime)Y  \bigr\}(1+ f_A(\mathbf p))(1+ f_A(\mathbf p^\prime)) -  \bigl\{ (1+f(\mathbf k)), \, Y (1+\bar f(\mathbf k^\prime))Y  \bigr\} f_A(\mathbf p) f_A(\mathbf p^\prime) \right] \notag   .
\end{align}
These expressions are strongly reminiscent of the corresponding single-flavor collision terms, except for their ``non-abelian'' structure: the distributions (i.e., density matrices) and scattering/annihilation matrix elements ($Y$) do not commute.  
The total collision term is 
\be
\label{totalcollision}
\mathscr{C}[f,\bar f] = g_* \bigl( \, \mathscr{C}[f,\bar f]_{\textrm{scat}} + \mathscr{C}[f,\bar f]_{\textrm{ann}} \, \bigr) \; ,
\ee
where we include the additional factor $g_* \sim 200$ to mimic the true number of degrees of freedom in the plasma during the EWPT.

There is one important subtlety: what is the dispersion relation for $k^0$ (and $k^{\prime0}$)?\footnote{This discussion applies to both particle $(k^0 >0$) and antiparticle ($k^0 < 0$) poles.  In evaluating $\mathscr{C}$, we have $k^0,k^{\prime 0} > 0$, where all negative energies are made positive through a change of variables.}
Working to linear order in $\epsilon$, it is consistent to set $k^0 \!=\! \bar{\omega}_{\mathbf k}$ in $\mathscr{C}$, since corrections will be $\mathcal{O}(\ecoll \eosc)$.
However, in that approximation, the collision term relaxes the density matrices to a ``false equilibrium'' $f, \bar{f} \!\to\! \textrm{diag} (n_B(\bar{\omega}_{\mathbf k}),n_B(\bar{\omega}_{\mathbf k}))$, rather than the ``true equilibrium'' $f, \bar{f} \!\to\! \textrm{diag}(n_B(\omega_{1\mathbf k}),n_B(\omega_{2\mathbf k}))$, with the correct dispersion relations.  This distinction is crucial: according to Eq.~\eqref{eq:qboltz}, the CP-violating source term vanishes if $f,\bar{f}$ are proportional to the identity, which they are when ``falsely'' equilibriated.  Hence, the $\mathcal{O}(\eosc^0)$ collision term quenches charge generation, an unphysical effect.

To remedy this issue, when evaluating the collision term we replace the lowest order solution to the constraint equations given in  Eq.~\ref{eq:treesolutionv1} with 
\begin{equation}
\label{eq:treesolutionv2}
\begin{split}
G_{ij}^{>}(k,x)  &= 2\pi\delta(k^2 - {m}_{ij}^2) \, \left[\, \theta(k^0)(\delta_{ij} + f_{ij}(\mathbf k, x)) + \theta(-k^0) \bar {f}_{ij}(-\mathbf k,x) \, \right] ~,  \\
G_{ij}^{<}(k,x)  &= 2\pi\delta(k^2 - {m}_{ij}^2) \, \left[\,\theta(k^0) f_{ij} (\mathbf k,x) + \theta(-k^0)(\delta_{ij}+ \bar f_{ij} (-\mathbf k,x)\, \right] \,.
\end{split}
\end{equation}
with $m_{ij}^2 = 1/2 (m_i^2 + m_j^2)$, implying
\be
k^0 = \omega^{ij}_{\mathbf k} \equiv \left\{\ba{lll} \omega^{i}_{\mathbf k} & \quad & i=j \\ \bar{\omega}_{\mathbf k} & \quad & i \ne j \\ \ea \right. \; .
\label{massresum}
\ee
This procedure can be viewed as the resummation of a class of $\mathcal{O}(\ecoll \times \eosc^n)$ corrections to the collision terms, corresponding to dynamical effects over length scales $L_{\textrm{mfp}}/\eosc^n$ ($n>1$).  It is physically justified to neglect these corrections for off-diagonal modes: these modes are damped to zero on a scale $L_{\textrm{mfp}}$ and are not sensitive to longer scales $L_{\textrm{mfp}}/\eosc^n$.  In contrast, diagonal modes do not equilibrate on a scale $L_{\textrm{mfp}}$; they approximately equilibrate to $f, \bar{f} \sim \textrm{diag} (n_B(\bar{\omega}_{\mathbf k}),n_B(\bar{\omega}_{\mathbf k}))$, but only reach true equilibrium over longer scales $L_{\textrm{mfp}}/\eosc^n$.  Therefore, by adopting Eq.~\eqref{massresum}, we are evaluating collision terms involving diagonal modes to all orders in $\eosc$, as is required to treat equilibration properly.  Progress in evaluating $\mathscr{C}$ to all orders in $\eosc$ for {\it both} diagonal and off-diagonal modes has been made in Ref.~\cite{Herranen:2008hi}.

With this prescription, a compact matrix expression for $\mathscr{C}$ is no longer possible and we must write all mass basis indices explicitly.  The collision terms are
\begin{align}
\label{collisionscat}
\mathscr{C}^{\textrm{scat}}_{ij}[f,\bar f] =& \int \!\! \frac{d^3 k^\prime}{(2\pi)^3} \; \left( \delta_{ia} Y_{bc} Y_{dj} + Y_{ic} Y_{da} \delta_{bj} \right) \\
& \times \left( R^{\textrm{scat,in}}_{abcd}(\mathbf k,\mathbf k^\prime) \, (1+f(\mathbf k))_{ab} \, f_{cd}(\mathbf k') - R^{\textrm{scat,out}}_{abcd}(\mathbf k,\mathbf k^\prime) \, f_{ab}(\mathbf k) \, (1+f(\mathbf k'))_{cd} \right) \notag \\
\label{collisionann}
\mathscr{C}^{\textrm{ann}}_{ij}[f,\bar f] =& \int \!\! \frac{d^3 k^\prime}{(2\pi)^3} \; \left( \delta_{ia} Y_{bc} Y_{dj} + Y_{ic} Y_{da} \delta_{bj} \right) \\
&\times \left(  R^{\textrm{ann,in}}_{abcd}(\mathbf k,\mathbf k^\prime) \, (1+f(\mathbf k))_{ab} \, (1+\bar f(\mathbf k^\prime))_{cd} - R^{\textrm{ann,out}}_{abcd}(\mathbf k,\mathbf k^\prime) \, f_{ab}(\mathbf k) \, \bar f_{cd}(\mathbf k^\prime) \right)   \notag
\end{align}
The scattering kernels are
\begin{align}
{R}^{\textrm{scat,in}}_{abcd}(\mathbf k,\mathbf k^\prime) & = 
\frac{T \, n_B(t_0)}{64\pi t \, \omega_{\mathbf{k}}^{ab} \, \omega_{\mathbf{k}^\prime}^{cd}} \, \theta(t^2-t_0^2)  \, \log\left( \frac{1+n_B(t_-)}{1+n_B(t_+)} \right)  \\
{R}^{\textrm{scat,out}}_{abcd}(\mathbf k,\mathbf k^\prime) & =  \frac{T (1+n_B(t_0))}{64\pi t \, \omega_{\mathbf{k}}^{ab} \, \omega_{\mathbf{k}^\prime}^{cd}} \, \theta(t^2-t_0^2)  \, \log\left( \frac{1+n_B(t_-)}{1+n_B(t_+)} \right) 
\end{align}
and the annihilation kernels are
\begin{align}
{R}^{\textrm{ann,in}}_{abcd}(\mathbf k,\mathbf k^\prime) & =  \frac{T \, n_B(s_0)}{128\pi s \, \omega^{ab}_{\mathbf k} \, \omega^{cd}_{\mathbf k^\prime}} \, \theta(s^2_0-s^2-4m_A^2)  \, \log\left( \frac{n_B(s_-)n_B(-s_-)}{n_B(s_+)n_B(-s_+)} \right)  \\
{R}^{\textrm{ann,out}}_{abcd}(\mathbf k,\mathbf k^\prime) & = \frac{T\, (1+n_B(s_0))}{128\pi s \, \omega^{ab}_{\mathbf k} \, \omega^{cd}_{\mathbf k^\prime}} \, \theta(s^2_0-s^2-4m_A^2)  \, \log\left( \frac{n_B(s_-)n_B(-s_-)}{n_B(s_+)n_B(-s_+)} \right) 
\end{align}
where 
\begin{align}
t &\equiv | \mathbf k - \mathbf k^\prime| , & t_0 &\equiv \omega^{ab}_{\mathbf k} - \omega^{cd}_{\mathbf{k}^\prime} , &
t_\pm &\equiv \, \pm \, \frac{t_0}{2} + \frac{t}{2} \, \sqrt{1+ 4 m_A^2/(t^2 - t_0^2) } \; ,  \\
s &\equiv | \mathbf k + \mathbf k^\prime| , & s_0 &\equiv \omega^{ab}_{\mathbf k} + \omega^{cd}_{\mathbf{k}^\prime} , &
s_\pm &\equiv \frac{s_0}{2} \pm \frac{s}{2} \, \sqrt{1+ 4 m_A^2/(s^2 - s_0^2) } \; .
\end{align}
From these expressions, one can verify several facts.  First, detailed balance is satisfied since
\be
{R}^{\textrm{scat,out}}_{abcd}(\mathbf k,\mathbf k^\prime) = e^{t_0/T} \, {R}^{\textrm{scat,in}}_{abcd}(\mathbf k,\mathbf k^\prime) \; , \qquad 
{R}^{\textrm{ann,out}}_{abcd}(\mathbf k,\mathbf k^\prime) = e^{s_0/T} \, {R}^{\textrm{ann,in}}_{abcd}(\mathbf k,\mathbf k^\prime) \; .
\ee
Second, $\mathscr{C}$ vanishes for
\be
f(\mathbf k) = \left( \ba{cc} n_B(\omega_{1 \mathbf k}- \mu_1) & 0 \\ 0 & n_B(\omega_{2 \mathbf k}-\mu_2) \ea \right) \; , \qquad
\bar{f}(\mathbf k) = \left( \ba{cc} n_B(\omega_{1 \mathbf k}+\mu_1) & 0 \\ 0 & n_B(\omega_{2 \mathbf k}+\mu_2) \ea \right) \; ,
\ee
with chemical potentials $\mu_{1,2}$.  (If $Y$ is diagonal in the mass basis, $\phi_{1,2}$ charges are separately conserved; otherwise only total charge $\phi_1 + \phi_2$ is conserved and $\mu_1=  \mu_2$.)  Third, the continuity equation is satisfied provided
$\textrm{Tr} \int d^3k/(2\pi)^3 (\mathscr{C}[f,\bar f] - \mathscr{C}[\bar f,f])=0$, which follows from the relations
\be
{R}^{\textrm{scat,out}}_{abcd}(\mathbf k,\mathbf k^\prime) = {R}^{\textrm{scat,in}}_{cdab}(\mathbf k^\prime,\mathbf k) ,\;\;
{R}^{\textrm{ann,in}}_{abcd}(\mathbf k,\mathbf k^\prime) = {R}^{\textrm{ann,in}}_{cdab}(\mathbf k^\prime,\mathbf k), \; \; 
(\textrm{in} \leftrightarrow \textrm{out}) \; .
\ee


\section{Decoupling in the collision term}
\label{app:decoupling}

In Ref.~\cite{konstandin}, the ansatz \eq{collisionansatz}, $\mathscr{C}_{12} = -\Gamma_{12} f_{12}$, was made for the off-diagonal collision term. Here we consider corrections to this ansatz and when it may be justified.

In general the collision terms given by \eqs{collisionscat}{collisionann} have much more complicated structure than this simple ansatz. We can simplify them somewhat by working to linear order in deviations from equilibrium, taking
\be
f_{ij} = f^{\text{eq}}_{ij} + n_B(\omega_{ij})(1+n_B(\omega_{ij}))\delta f_{{ij}}\,,
\ee
where $f^{\text{eq}}_{ij} = n_B(\omega_i)\delta_{ij}$, and $\delta f_{ij}\sim \epsilon$ for some small $\epsilon$. We have factored out $n_B(1+n_B)$ in the $\mathcal{O}(\epsilon)$ term for later notational convenience (cf. \cite{Arnold:2003zc}). Then, we linearize the collision terms in $\delta f$. At $\mathcal{O}(\epsilon^0)$, the collision terms vanish (as required for the equilibrium distributions). Now, consider  the off-diagonal collision terms $\mathscr{C}_{12}$ linearized in $\delta f$. (We will study the structure of the diagonal collision terms in future work.) It can be organized into three sets of terms,
\be
\mathscr{C}_{12} = \mathscr{C}_{12}^{\text{loss}} + \mathscr{C}_{12}^{\text{gain}} + \mathscr{C}_{12}^{\text{source}}\,.
\ee
The loss term  takes the form of the ansatz \eq{collisionansatz} but with a $\vect{k}$ dependent relaxation rate,
\be
\label{lossterm}
\begin{split}
\mathscr{C}_{12}^{\text{loss}}(\vect{k},z) &= -\Gamma_{12}(\vect{k},z)  f_{12}(\vect{k},z) = -\Gamma_{12}(\vect{k},z)  n_B(\omega_{12})(1+n_B(\omega_{12}))\delta f_{12}(\vect{k},z)\,,
 \end{split}
\ee
where
\be
\label{relaxationrate}
\begin{split}
\Gamma_{12}(\vect{k},z) &= \frac{g_*}{1+ n_B(\omega_{12})}\int \frac{d^3k' }{(2\pi)^3}\\ 
&\quad \times \! \biggl\{\! (Y_{11}^2 + Y_{12}Y_{21}) \bigl[ R_{1211}^{\text{scat,out}}(\vect{k},\vect{k}')(1+n_B(\omega_{\vect{k}'}^1 \!)) + R_{1211}^{\text{ann,out}}(\vect{k},\vect{k}')n_B(\omega_{\vect{k}'}^1 \!)\bigr] \\
&\qquad + (Y_{22}^2 + Y_{12}Y_{21}) \bigl[ R_{1222}^{\text{scat,out}}(\vect{k},\vect{k}')(1+n_B(\omega_{\vect{k}'}^2\!)) + R_{1222}^{\text{ann,out}}(\vect{k},\vect{k}')n_B(\omega_{\vect{k}'}^2\!)\bigr]\! \biggr\}.
\end{split}
\ee
Meanwhile, the ``gain'' term is
\be
\label{gainterm}
\begin{split}
\mathscr{C}^{\text{gain}}_{12}(\vect{k},z) &= g_* \int \frac{d^3k' }{(2\pi)^3} \sum_{i=1}^2\\ 
&\quad \times\Bigl\{  R_{ii12}^{\text{scat,out}}(\vect{k},\vect{k}')n_B(\omega_{\vect{k}}^i)[1+n_B(\omega_{\vect{k}'}^{12})][Y_{11}Y_{22}\delta f_{12}(\vect{k}') + Y_{12}^2 \delta f_{21}(\vect{k}')] \\
&\qquad -R_{ii12}^{\text{ann,out}}(\vect{k},\vect{k}')n_B(\omega_{\vect{k}}^i)n_B(\omega_{\vect{k}'}^{12})[Y_{11}Y_{22}\delta \bar f_{12}(\vect{k}') + Y_{12}^2 \delta \bar f_{21}(\vect{k}')] \Bigr\}\,,
\end{split}
\ee
and the ``source'' term is
\be
\begin{split}
\label{collisionalsource}
\mathscr{C}^{\text{source}}_{12}(\vect{k},z) &= g_*Y_{12} \int \frac{d^3k' }{(2\pi)^3}\sum_{i,j=1}^2 Y_{jj} \\
&\quad\times   \Bigl\{ R_{iijj}^{\text{scat,out}}(\vect{k},\vect{k}')n_B(\omega_{\vect{k}}^i)[1+n_B(\omega_{\vect{k}'}^j)][\delta f_{jj}(\vect{k}') - \delta f_{ii}(\vect{k})] \\
&\qquad - R_{iijj}^{\text{ann,out}}(\vect{k},\vect{k}')n_B(\omega_{\vect{k}}^i)n_B(\omega_{\vect{k}'}^j)[\delta \bar f_{jj}(\vect{k}') + \delta f_{ii}(\vect{k})]\Bigr\}\,,
\end{split}
\ee
so named since deviations of diagonal distributions of equilibrium act as a source for $f_{12}$ through a nonzero off-diagonal coupling $Y_{12}$.

The ansatz \eq{collisionansatz} misses the effects of both the collisional gain and source terms. In the limit of flavor-blind interactions, $y_L = y_R \equiv y$, we have $Y_{12}=0$ and $Y_{11,22} = y$, so the source terms vanish. However, a part of the gain term still remains. The remaining terms (so-called ``noise terms'' in \cite{Cline:2000nw}) are often neglected by assuming $\delta f(\vect{k}')$ to be a randomly fluctuating variable, causing the integral over $\vect{k}'$ to be suppressed relative to the loss term. Here a similar suppression may happen because of oscillations of $\delta f_{12}$ in both $\vect{k'}$ and $z$ with frequency $\omega_1-\omega_2$, but there is otherwise no a priori reason to drop these terms.

In the comparison shown in Fig.~\ref{fig:comp2} which uses  the baseline parameters of Table~\ref{tab:baseline}, it is the case that $Y_{12} < Y_{11,22}$, but still nonzero. Thus the approximate solution using the ansatz \eq{collisionansatz} for the collision term misses the collision-induced source \eq{collisionalsource} for $f_{12}$. This is one likely cause of the smaller normalization of the approximate solution even in the adiabatic regime where \eq{decoupledf12solution} is otherwise  valid.


\section{Power Counting the Off-Diagonal Distribution}
\label{app:power}

In Sec.~\ref{ssec:offdiagonal}, we argued that the $f_{12}$ in \eq{decoupledf12solution} is actually $\mathcal{O}(\ewall^0)$, not $\mathcal{O}(\ewall)$ as argued in \cite{konstandin}, and therefore could not be neglected in the source term \eq{eq:matrix-coupling}.  However, since in our power counting
$f_{12}$ is still $\mathcal{O}(\eosc)$ (see \eq{S12def}), it is fair to ask why we still do not neglect the terms containing $f_{12,21}$ in \eq{eq:matrix-coupling} since they become $\mathcal{O}(\epsilon^2)$. The reason is that, by power counting $f_{ij}$ this way,  \emph{every} component of the source term \eq{eq:matrix-coupling} becomes $\mathcal{O}(\epsilon^2)$. The diagonal and off-diagonal components are all the same order and should all be kept. Counting consistently, the leading nontrivial terms in the kinetic equation become $\mathcal{O}(\epsilon^2)$. Deviations of $f$ from equilibrium can be counted as one power of $\epsilon$. Then the oscillation term is $\eosc\epsilon$, the source term is $\ewall\epsilon$, and the collision term is $\epsilon_{\rm int}\epsilon$. So the derivative $\partial_z$ on the left-hand side of the kinetic equation always brings down at least one $\epsilon$ when acting on $f$, and the whole kinetic equation begins (nontrivially) at $\mathcal{O}(\epsilon^2)$. The upshot is that, counting consistently at this order, the full coupled matrix structure of the kinetic equation must be kept.

This exercise in power counting also tells us that there \emph{is} a regime in which the solution \eq{decoupledf12solution} from Ref.~\cite{konstandin} is a relatively good approximation for $f_{12}$. In the adiabatic regime, when $L_{w}\gg L_{\rm osc}$, the  factor $\exp(-i\Delta\omega z'')$  cuts off the range of integration to be of order $L_{\rm osc}$ (over a larger region,  oscillations average $f_{12}$ out to zero). Then $f_{12}$ in \eq{decoupledf12solution} is order
\be
\label{f12adiabaticestimate}
f_{12} \sim (\text{range of integration})\times \Sigma_{12}\times [n_B(\omega_1) - n_B(\omega_2)] \sim \frac{1}{\eosc}\times \ewall\times \eosc = \ewall\,
\ee
which \emph{is} consistent with Ref.~\cite{konstandin}. 
This is just the part of $f_{12}$ sourced by the equilibrium diagonal distributions (call it $f_{12}[n_B]$). One can show the deviations $\delta f_{11,22}$ of the diagonal distributions from equilibrium sourced by $f_{12}[n_B]$ is then order $\ewall(\ewall/\eosc)$, which then feeds back to source an additional part of $ f_{12}$ (call it $f_{12}[\delta f]$) of order $\ewall(\ewall/\eosc)^2$.
Since $\ewall\ll \eosc$ in the adiabatic regime, these additional deviations are suppressed relative to $f_{12}[n_B]$. So \eq{decoupledf12solution} should be a good approximation for $f_{12}$ in the adiabatic regime.
However, as  explained in Sec.~\ref{ssec:offdiagonal}, one should {\it not} decouple $f_{12}$ from the evolution of the diagonal densities in the source  \eq{eq:matrix-coupling}, even though $\delta f_{11,22}$ are formally $\mathcal{O}(\epsilon_{\textrm{wall}}^2)$ in this regime.  Neglecting this source, and thereby neglecting deviations from equilibrium in $f_{ii}, \bar{f}_{ii}$, precludes the existence of diffusion and therefore grossly underestimates the total charge in the unbroken phase.

\end{document}